\documentclass[useAMS,usenatbib]{mn2e}
\usepackage{graphicx}

\newif\ifAMStwofonts

\title[]
{Near Infrared properties of 12 Globular Clusters toward the inner Bulge of the Galaxy \thanks{Based on data taken at the ESO/NTT Telescope within the observing programs 079.D-0200 and 081.D-0371 }}
\author[Valenti, Ferraro \& Origlia]
       {E. Valenti$^{1,2}$, F.~R. Ferraro$^3$ \& L. Origlia$^4$\\
        $^1$ESO \-- European Southern Observatory, Casilla 19001 Santiago, Chile; evalenti@eso.org\\
        $^2$ Pontificia Universidad Catolica de Chile, Dep. de Astronomia y Astrofisica, , Santigo, CHILE\\
        $^3$Bologna University, Astronomy Department, Bologna, Italy\\
        $^4$INAF \-- Osservatorio Astronomico di Bologna, Italy
        }
\date{}

\pagerange{\pageref{firstpage}--\pageref{lastpage}}
\pubyear{2008}

\begin{document}

\maketitle

\label{firstpage}

\begin{abstract}

We present near\--IR Colour\--Magnitude diagrams and physical parameters for a sample of 12 galactic globular clusters located toward the inner Bulge region.
For each cluster we provide measurements of the reddening, distance, photometric metallicity,  luminosity of the horizontal branch red clump, and of the red giant branch bump and tip. The sample discussed here together with that presented in \citet{io07} represent the largest homogeneous catalog of Bulge globular clusters (comprising $\sim80\%$ of the entire Bulge cluster population) ever studied. 

The compilation is available in electronic form on the World Wide Web ({\tt http://www.bo.astro.it/$\sim$GC/ir\_archive})
\end{abstract}


\section{Introduction}
Galactic globular clusters (GCs) play a crucial role in stellar astrophysics and cosmology. Indeed, the detailed study of their stellar content addresses important questions ranging from stellar structure, evolution, dynamics, to Galaxy formation and the early epoch of chemical enrichment.
Being the largest aggregates in which all the post Main Sequence stars can be individually resolved, they also represent fiducial templates for understanding the integrated light from distant unresolved stellar systems. 

In particular, the GCs in the Bulge provides ideal templates for exploring the high metallicity domain, and for studying the stellar content of extragalactic bulges and ellipticals.
However, the high and variable foreground extinction, the severe crowding (blending) and the modest performances of the past generation of near\--IR instrumentation, prevented detailed and systematic surveys  of most of the cluster population toward the Bulge.

Only few IR studies have been devoted to explore some of the GCs in the inner spheroid \citep[see][and references therein]{min95,dav00,dav01}, providing reddening, distance and photometric metallicity estimates.

Trying to supply the lack of complete and homogeneous survey, our group started a long\--term project devoted to fully characterize  the entire cluster populations in the Bulge using colour\--magnitude diagrams (CMDs) and luminosity functions (LFs) in the near\--IR (see \citet[][hereafter F00]{frf00}; \citet[][hereafter VFO04a]{io04a}; \citet[][hereafter VFO04b]{io04b}; \citet{io04}; \citet{io05}; \citet{ori05}). 
It is worth to mention that, the targets selection was performed by defining "{\it Bulge GCs}" as all those located within 3~Kpc of the Galactic center, and  within $|b|\leq10^{\circ}$ and $|l|\leq20^{\circ}$, where l and b are the Galactic coordinates. This definition is mainly a working hypothesis and refers only to the spatial distribution of the clusters within the Bulge.  
The collected photometric database has been used to perform a detailed description of the main morphological and evolutionary features of the red giant branch (RGB) sequence, by means of a set of photometric indices that were defined and widely described in F00 and calibrated as a function of the cluster metallicity in VFO04a and VFO04b. In \citet[][hereafter {\it Paper~I}]{io07} we studied 24 Bulge GCs providing: {\it i)} the photometric catalog; {\it ii)} the RGB mean ridgeline; {\it ii)} accurate reddening, distance and metallicity determinations; {\it iv)} the luminosity of the main RGB evolutionary features (e.g. the bump and the tip); and {\it v)} the magnitude of the horizontal brach red clump.  It is worth mentioning that the cluster properties were derived in a self\--consistent way. In fact, the derived cluster reddenings, distances, and metallicities are based on (1) a homogenous photometric database analyzed using the same data reduction procedures, and calibrated onto the 2MASS photometric system; (2) the \citet{frf99} distance scale; and (3) a uniform and high\--resolution metallicity scale \citep{cg97}.
To be easily accessible by the community, the compilation was also provided in electronic form at the following WEB address: {\tt http://www.bo.astro.it/$\sim$GC/ir\_archive}.

\begin{figure}
\includegraphics[width=8cm]{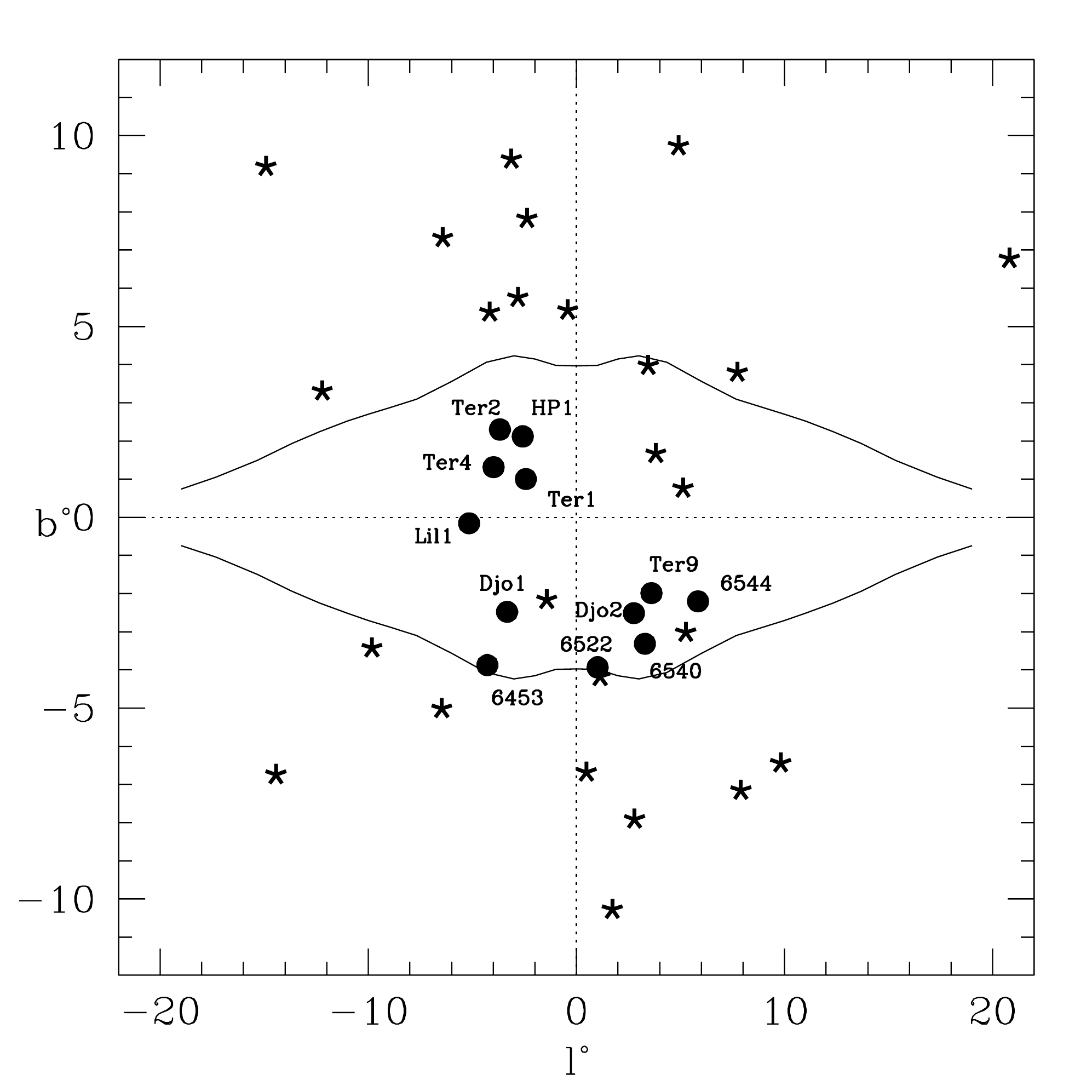}
\caption{Position of the Bulge cluster global sample with respect to the COBE/DIRBE 3.5$\mu$m inner bulge outline (solid line; \citet{wei94} at 5~MJy~sr$^{-1}$). Filled circles are the clusters presented here; clusters published in {\it Paper~I} are marked with star symbols.}
\label{map}
\end{figure}

This paper deals with an additional sample of 12 GCs located in the innermost region of the Bulge (see Fig.~\ref{map}).
In a forthcoming paper we will then present  the last sample of 10 clusters towards the outer Bulge direction.

In \S~2 we described the observations and data reduction procedure, while in \S~3 we briefly summarize the method used to derived the cluster reddening, distance, metallicity, and the RGB evolutionary features. The derived clusters properties are presented in \S~4, and discussed in \S~5.

\section{Observations and data reduction}

A set of J, H and K images of 12 Bulge clusters, namely Ter~1, Ter~2, Ter~4, Ter~9, Lill~1, HP~1, Djo~1, Djo~2, NGC~6540, NGC~6544, NGC~6522, NGC~6453, was obtained at the European Southern Observatory (ESO), La Silla during two observing runs in July 2007 and June 2008. We used the near\--IR camera SofI, mounted at the ESO New Technology Telescope (NTT)  in large field\--mode which is characterized by a pixel size of 0.288" and a total field of view of 4.9'$\times$4.9'. During the two runs both {\it long} and {\it short} exposures were secured for each cluster, the latter to avoid the saturation of bright giants. The {\it long} exposures are a combination of 45, 75 and 100 sub\--exposures each 3~sec long, in J, H and K$_s$ respectively. The {\it short} images have been obtained with a combination of 30 exposures, each 1.2~sec long, in all the pass\--bands. All the secured images are roughly centered on the cluster center
covering a region which allow us to sample a significant fraction of the total cluster light (typically $\sim$ 80\% -- 95\%) in all the program clusters. During the observations the average seeing was always quite good (FWHM$\approx$0.8" \--1"). Every image has been background\--subtracted using sky fields located several arcminutes away from the cluster center and flat\--field corrected using halogen lamp exposures, acquired with the standard SofI calibration setup.

Standard crowed\--field photometry, including point\--spread function modeling, was carried out on each frame using DAOPHOT~II ALLSTAR \citep{dao}. For each cluster, two photometric catalogs (derived from the long and short exposures) listing the instrumental J, H and K$_s$ magnitudes were obtained by cross\--correlating the single\--band catalogs. The short and long catalogs were combined by means of a proper weighted average, weighting the short exposures of brightest stars higher. The internal photometric accuracy was estimated from the rms frame\--to\--frame scatter of multiple star measurements. Over most of the RGB extension, the internal errors are quite low ($\sigma_J\sim\sigma_H\sim\sigma_K < $0.03 mag), increasing up to $\sim 0.06$ mag at $K_s\geq16$.

The instrumental magnitudes were then converted into the Two Micron All Sky Survey (2MASS) photometric system\footnote{An overall uncertainty of $\pm0.05$ mag in the zero\--point calibration in all the three bands has been estimated.}, and the star positions astrometrized onto 2MASS\footnote{The astrometric procedure provided rms residuals of $\approx$0.2" both right ascension and declination.} as done for the sample published in {\it Paper~I}.

\begin{figure*}
\includegraphics[width=8.5cm, height=9.5cm]{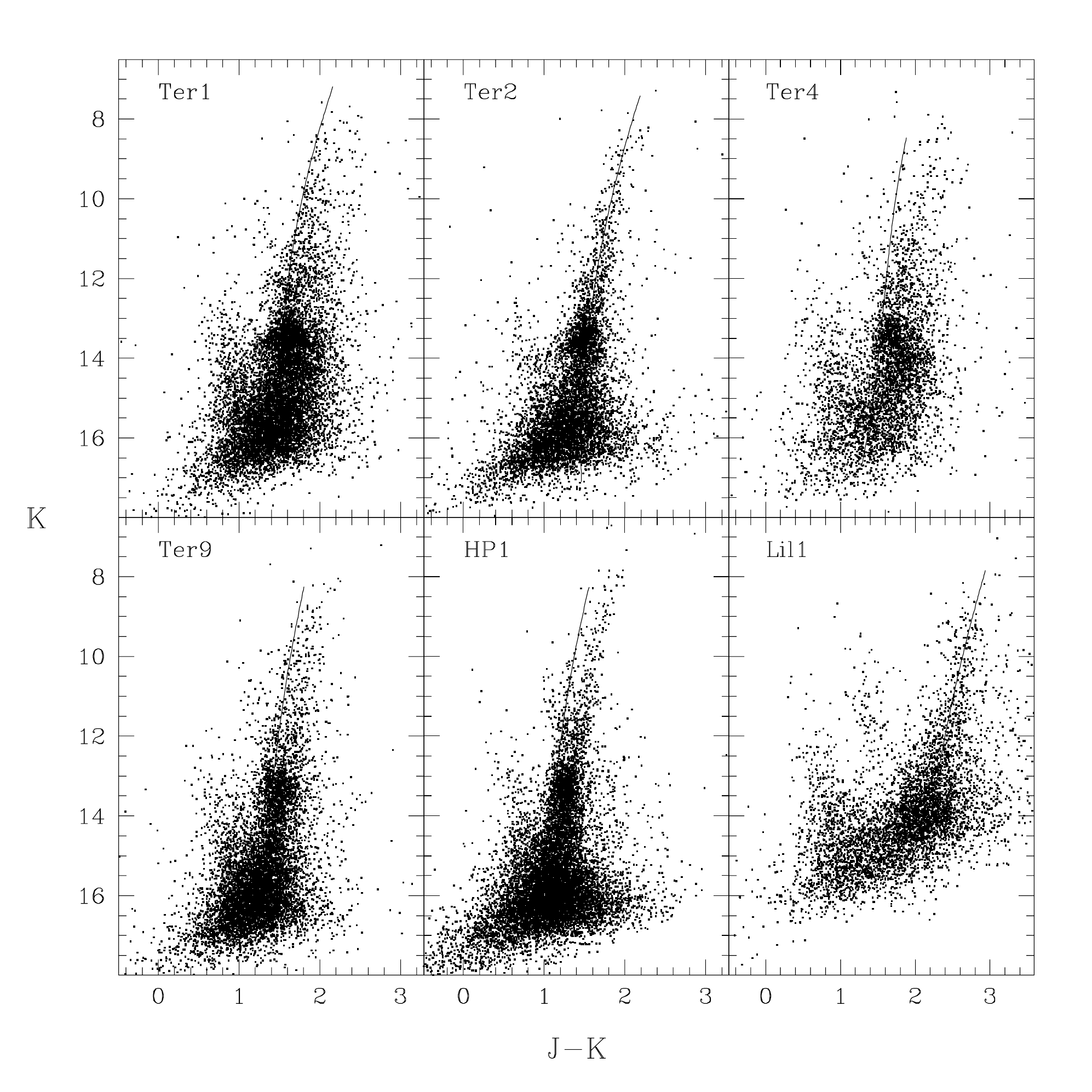}
\includegraphics[width=8.5cm, height=9.5cm]{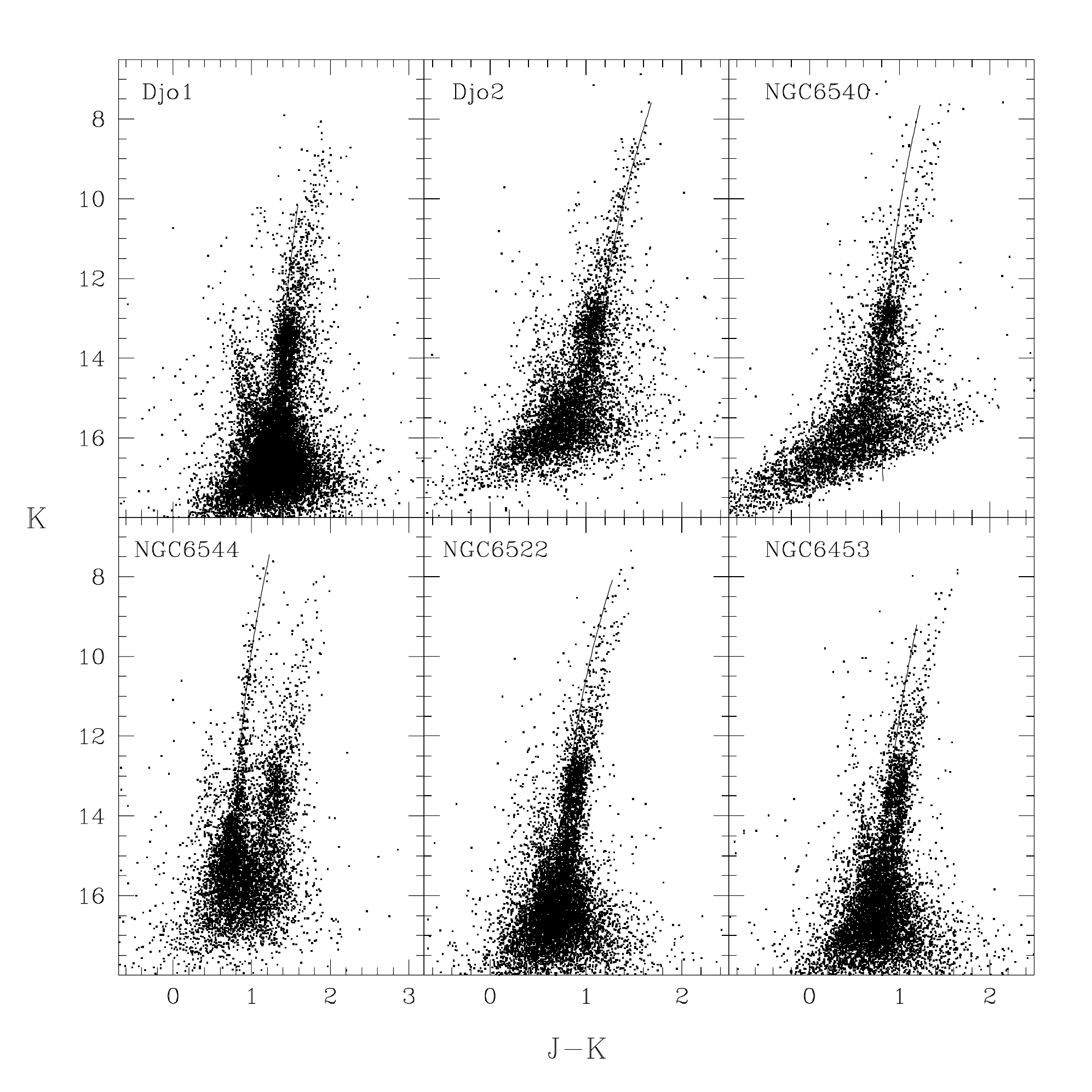}
\caption{Observed near\--IR CMDs and derived RGB ridgelines for the program clusters}
\label{cmds}
\end{figure*}

Figure~\ref{cmds} shows the derived IR CMDs for this new sample of 12 clusters. The photometric catalogs span the entire RGB extension, from the tip to $\approx$2-5 mag below the HB (depending on the cluster extinction and with the possible exception of Lil~1) thus allowing a detailed analysis of the morphological and evolutionary properties of the clusters.

\section{Photometric Analysis}

The main cluster properties (e.g. reddening and distance) have been derived using a differential technique based on the comparison of CMDs and LFs of clusters with similar HB morphology.
This method allows one to derive reliable estimates of reddening
and distance only for intermediate\-- to high-metallicity
clusters, whose CMDs show a red clumpy HB morphology. In
fact, the location in magnitude of the HB LF peak can be used as
a reference feature and safely compared with that of the template
cluster (e.g. 47~Tuc). 
However, in 8 metal\--poor clusters (Ter~4, Ter~9, HP~1, Djo~1, NGC~6540, NGC~6544, NGC~6522 and NGC~6453) the blue HB morphology coupled with the relatively high reddening and remarkable level of field contamination prevented a reliable determination of the HB level, and so the use of the differential method. For these clusters the reddening and distance were derived using the empirical method presented by \citet[][hereafter FVO06]{fvo06}, which allows to simultaneously get the reddening, distance and metallicity of a stellar system using the RGB slope, tip, and the mean ridgeline of the [K, J--K] CMD. In order to derive the RGB fiducial ridgeline we used a low\--order polynomial to fit the RGB stars in the inner cluster region (see \S~4), rejecting those at ${\geq \pm 2 \sigma}$ from the best fit line.

The formal error on each derived quantity has been estimated by using a simple Monte Carlo simulation. In doing this, we randomly extracted a large number of values for each of the main observables (namely slope${JK}_{RGB}$ and K$^{tip}_{obs}$) from a Gaussian distribution peaked on the observed value, with $\sigma$ equal to the uncertainty of each observable. These values are used (following the FVO06 prescription) to derive metallicity, reddening and distance. The standard deviation of each set of values is the error associated to that specific quantity. However, beside the formal error obtained from this procedure [typically $\delta(m\--M)_0=\pm0.10 \-- 0.15$ mag, $\delta E(B\--V)=\pm0.03 \--0.04$ mag and $\delta [Fe/H]=\pm0.15$ dex] we estimate that conservative uncertainties for the derived quantities are $\delta (m\--M)_0=\pm0.2$ mag, $\delta E(B\--V)=\pm0.05$ mag, and  $\delta[Fe/H]=\pm0.2$ dex.

The derived cluster reddening and distance were used to
transform the observed CMDs and RGB ridgelines into the absolute
plane, and to measure the following parameters: (1) the
$(J\--K)_0$ and $(J\--H )_0$ colors at four fixed absolute magnitude
levels ($M_K = M_H =$ -5.5; -5; -4; and -3); (2) the absolute $M_K$
and $M_H$ magnitudes at constant $(J\--K)_0=(J\--H)_0=0.7$ colors;
and (3) the RGB slope in the [K, J\---K] - and [H, J\---H]-planes (see Tables \ref{jkRGB} and \ref{jhRGB}). Then,
using the empirical calibrations from VFO04a linking this set of
photometric indices to the cluster metal content,we finally derived
the photometric metallicity estimates in the \citet{cg97} scale. 
Hence, hereafter the notation [Fe/H] refers to the
\citet{cg97} scale. Note that to derive the global
metallicity [M/H], which takes into account the iron as well as
the ${\alpha}$\--element abundances, we used the calibrations presented
in FVO06 within the bulgelike enrichment scenario.
According to the more recent results on Bulge populations 
[\citet{lil1_B,ter4_B,ori08,zoc04,eug07,hp1_A,alv06} for Bulge clusters, and \citet{ro05,katia06,zoc06,lec07,fmr07,rov07} for giant field stars], we adopted $[\alpha/Fe]=0.30$
constant over the entire range of metallicity up to solar [Fe/H]. 

The observed CMDs and LFs have been also used to measure the location of the main RGB evolutionary features, such the bump and the tip. In doing this, we adopted the same strategy followed in VFO04b, in which the reader can find a detailed description of the procedures.
Finally, as done in {\it Paper~I} we used the bolometric corrections for Population~II giants computed by \citet{bol98} to transform the magnitude of the RGB tip and bump from the observed to the theoretical plane.

\section{The cluster properties}
In this section we present the CMDs and RGB fiducial ridgeline for the sample of 12 clusters. We briefly discuss the main CMD properties, and we provide some references for previous published photometry.
In Table~\ref{param} we listed the derived parameters for the cluster sample, namely the reddening, distance, and metallicity in both the adopted scales (cols. [9], [10], [11], and [12], respectively).
As done in {\it Paper~I}, using our distance estimates and the (l, b) Galactic coordinates from \citet{har}, and assuming a distance R$_0$=8 kpc to the Galactic center \citep{eis03}, we also provide new estimates of the clusters' distances from the Sun (d$_{\odot}$) and from the Galactic center (R$_{GC}$), and the distance components X, Y, Z in a Sun\--centered coordinate system (cols. [4], [5], [6], [7], and [8], respectively). Finally, Table~\ref{bol} lists the observed J, H and K magnitude of the HB red clump (cols. [4], [5], and [6]), and the J, H, K, and bolometric bump (cols. [7], [8], [9], and [13], respectively) and tip (cols. [10], [11], [12], and [14], respectively) magnitudes.

\subsection{Terzan~1}

\begin{figure} 
\includegraphics[width=8cm]{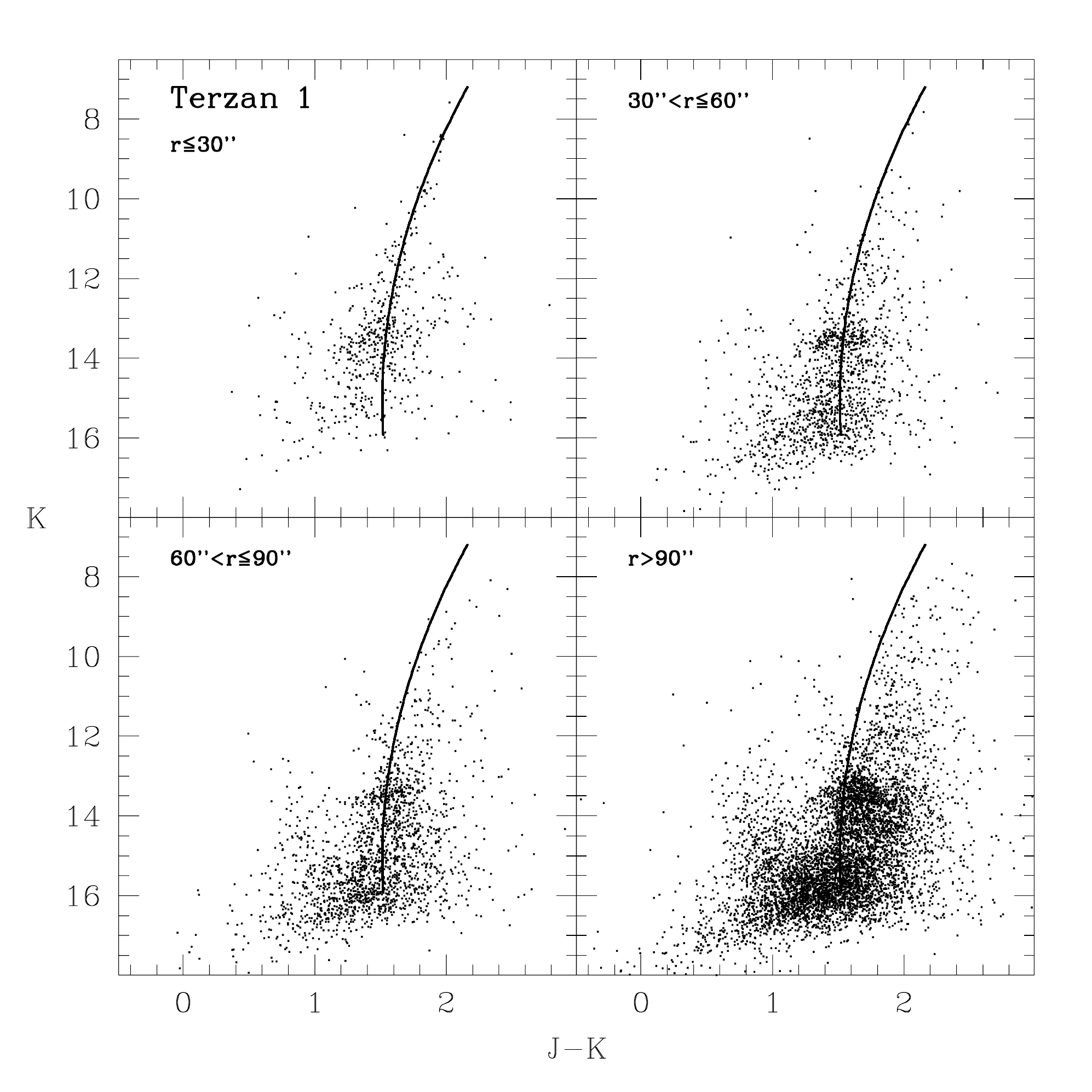}
\caption{[K, J--K] CMD of Terzan~1 at different distances (r) from the cluster center.}
\label{ter1}
\end{figure}

Terzan~1 is the most internal Bulge cluster of the observed sample (see Fig.~\ref{map}). Its reddening estimates found in literature range from E(B--V)=1.49 \citep[][hereafter H96]{har} to E(B--V)=2.48 \citep{ter1_B}. The only available photometric study is the one by \citet{ter1_B}, which provided [I, V--I] CMD based on HST/WPC2 data. The authors estimated the cluster reddening [E(B--V)=2.48] by comparing the derived CMD with that of NGC~6553, and the cluster distance [(m--M)$_0$=13.58] using the absolute V magnitude of the HB level. Because of the presence in their CMDs of a red HB clump coupled with a steep RGB, the author claimed that Ter~1 is one of the few (together with NGC~6388 and NGC~6441) second parameter cluster in the Galactic bulge.
The near\--IR CMD obtained here is shown in Fig.~\ref{cmds}, along with the derived RGB ridgeline. The most interesting features are {\it i)} a red clumpy HB at 1.4$\leq$(J\--K)${\leq}$0.8 and K$\sim$13.5;  {\it ii)} a large scattered RGB,  which appears to be caused by differential reddening and a high level of field contamination. The CMD, at different distances from the cluster center (see Fig.~\ref{ter1}) show that, in the innermost region (r$\leq30\arcsec$) , the RGB is quite narrow and the HB is red , while for increasing distances from the cluster center, the CMDs show a RGB significantly redder than the cluster mean loci. 
Moreover, at distances larger than 90$\arcsec$ from the center, a second clump at K$\sim$14.5 starts to be visible. This feature is likely due to field contamination, which is also responsable for the spread in the RGB. Considering only the stars within the first 30$\arcsec$ from the center and comparing the derived CMD and LF with those of 47~Tuc, we estimated a reddening of E(B\--V)=1.99 and a distance of (m\--M)$_0$=14.13. The photometric indices measured along the RGB in the [$M_K$, (J\--K)$_0$] and [M$_H$, (J\--H)$_0$] absolute planes (see also Tables~\ref{jkRGB} and \ref{jhRGB}), give the following metallicity estimates: [Fe/H]= --1.11~dex and [M/H]= --0.91~dex. Our metallicity estimates is in fair agreement with that of \citet{ter1_A}, who find a [Fe/H]=--1.3~dex from low resolution optical spectroscopy. The measured reddening is in between the range of values found in literature, while the distance derived from our IR photometry is significantly larger than the value found by  \citet{ter1_B}. However, when we take into account the $\Delta$E(B\--V)=0.5 the two estimates 
agree within the errors ($\sim$0.2mag).

\subsection{Terzan~2}
Terzan~2 has been observed in the near\--IR by \citet{ter2_B}, however the derived CMD is very shallow, sampling only very fews RGB stars.  From the analysis of the cluster RGB and assuming a metallicity of [Fe/H]$\sim$--0.25 \citep{az88}, the authors found a (m--M)$_0$=15 distance modulus and
a reddening of E(B--V)=1.25. Later on, \citet{ter2_A} presented a deeper optical photometry of Ter~2 from which they derived E(B--V)=1.54 and (m--M)$_0$=14.44. From the HB and RGB morphology, they concluded that the cluster should have a metallicity in between 47~Tuc (--0.7~dex) and NGC~6356 (--0.4~dex).

Our IR CMD (see Fig.~\ref{cmds}) shows a well defined red HB at K$\sim$13.6, suggesting a moderately high metallicity, like 47~Tuc and a quite narrow RGB. The analysis of the cluster CMDs and LFs yields a reddening slightly higher than the previous studies [E(B--V)=1.87], an intrinsic distance modulus (m--M)$_0$=14.35 consistent with the optical study, and metallicities [Fe/H]=--0.72~dex and [M/H]=--0.53~dex which are in good agreement with the value published by \citet{ter4lil1ter2} ([Fe/H]=--0.82~dex) based on medium resolution IR spectroscopy.

\subsection{Terzan~4}

\begin{figure} 
\includegraphics[width=8cm]{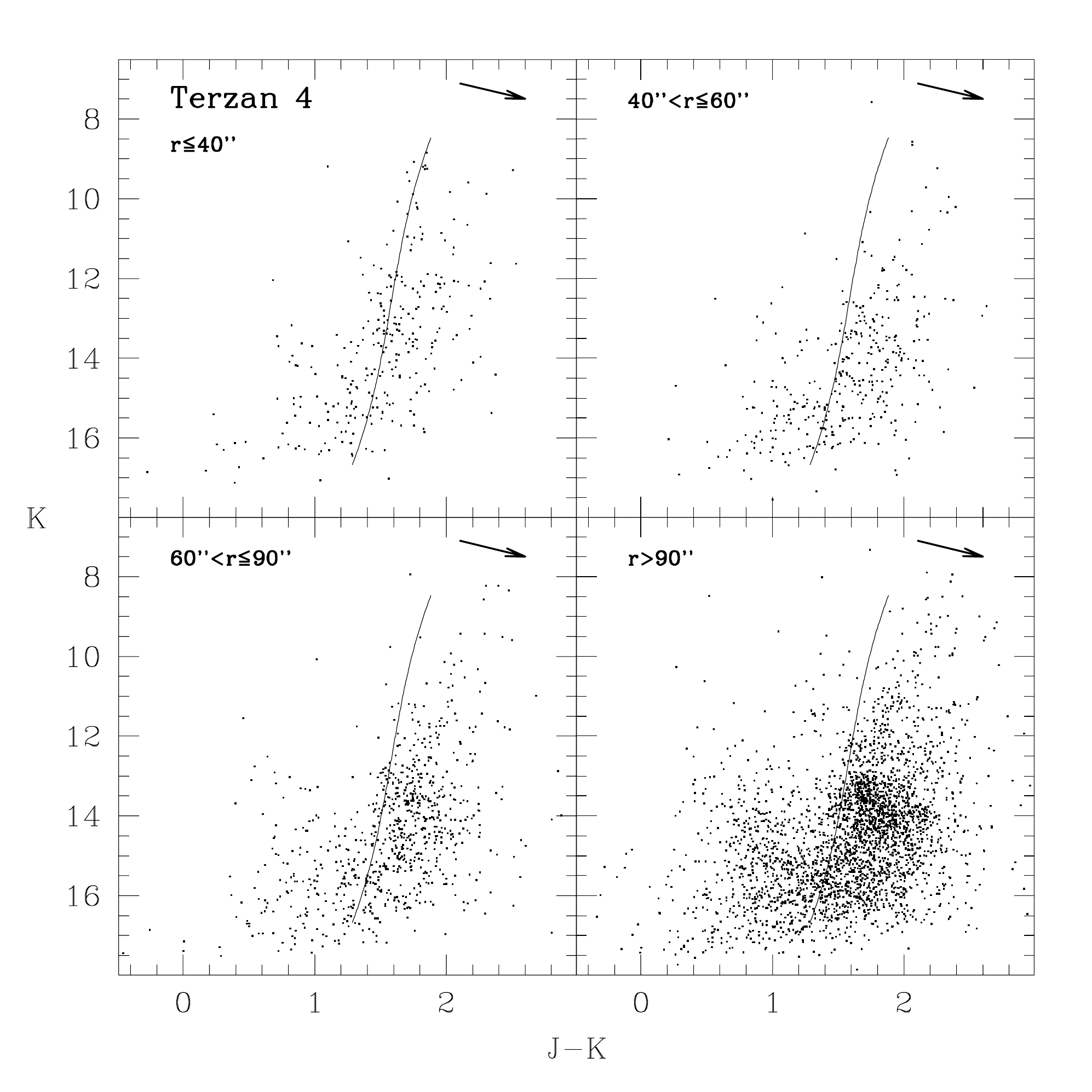}
\caption{[K, J--K] CMD of Terzan~4 at different distances (r) from the cluster center. The arrow represents the reddening vector.}
\label{ter4}
\end{figure}

Terzan~4 is a heavily reddened cluster which has been photometrically observed only in optical by \citet{ter4_C}, who presented ground\--based [V, V\---I] CMD. The authors derived estimates of the cluster reddening [E(B\--V)=2.35], distance [(m\--M)$_0$=14.59] and based on the similarity of the  [V, V\---I] CMD of Ter~4 and M~30, along with the presence of a blue HB, they suggested a low metallicity ([Fe/H]=\--2~dex). As shown in Fig.~\ref{cmds}, the IR CMD of Ter~4 is 
characterized by a spread bent RGB, a red clumpy HB and the disk main sequence at 10$\leq$K$\leq$17, and 0.5$\leq$(J\--K)$\leq$1.2. When looking at the CMD obtained considering only stars within 40$\arcsec$ from the cluster center (see Fig.~\ref{ter4}), the RGB appears quite steep suggesting a low\--intermediate metallicity, and a blue HB is barely visible at 15.5$\leq$K$\leq$13 and 0.8$\leq$(J\--K)$\leq$1.6. With increasing distance from the cluster center, the RGB gets progressively redder and more bent, and a red tilted HB becomes more populated. We derived the RGB ridge line using only the most internal region (r$\leq$40$\arcsec$). As shown in Fig.~\ref{ter4}, the RGB slope of the cluster is significantly steeper than the one at larger distances from the center. This suggests that the red bent RGB and the clumpy HB belong to the field population, which turns out to be more metal\--rich than the cluster itself. Moreover, the tilt in the red HB 
is likely due to differential reddening and to distance dispersion effect.
Since the cluster HB morphology and the high level of field contamination make difficult to derive the cluster reddening and distance by using the differential method, we adopted the empirical one (see FVO06). From the cluster CMD we estimated the RGB slope and the observed RGB tip (RGB$_{slope}$= --0.043 and K$^{tip}$=8.843, respectively). Adopting these values as input parameters, the computational routine (assuming the bulgelike scenario) gives a reddening E(B\--V)=2.05, an intrinsic distance modulus (m\--M)$_0$=14.13, and metallicity [Fe/H]=--1.58~dex and [M/H]=--1.37~dex.
Our metallicity estimate is in excellent agreement to the values published by \citet{ter4_B} ([Fe/H]=--1.6~dex) based on high resolution IR spectroscopy.

\subsection{Terzan~9}

\begin{figure} 
\includegraphics[width=8cm]{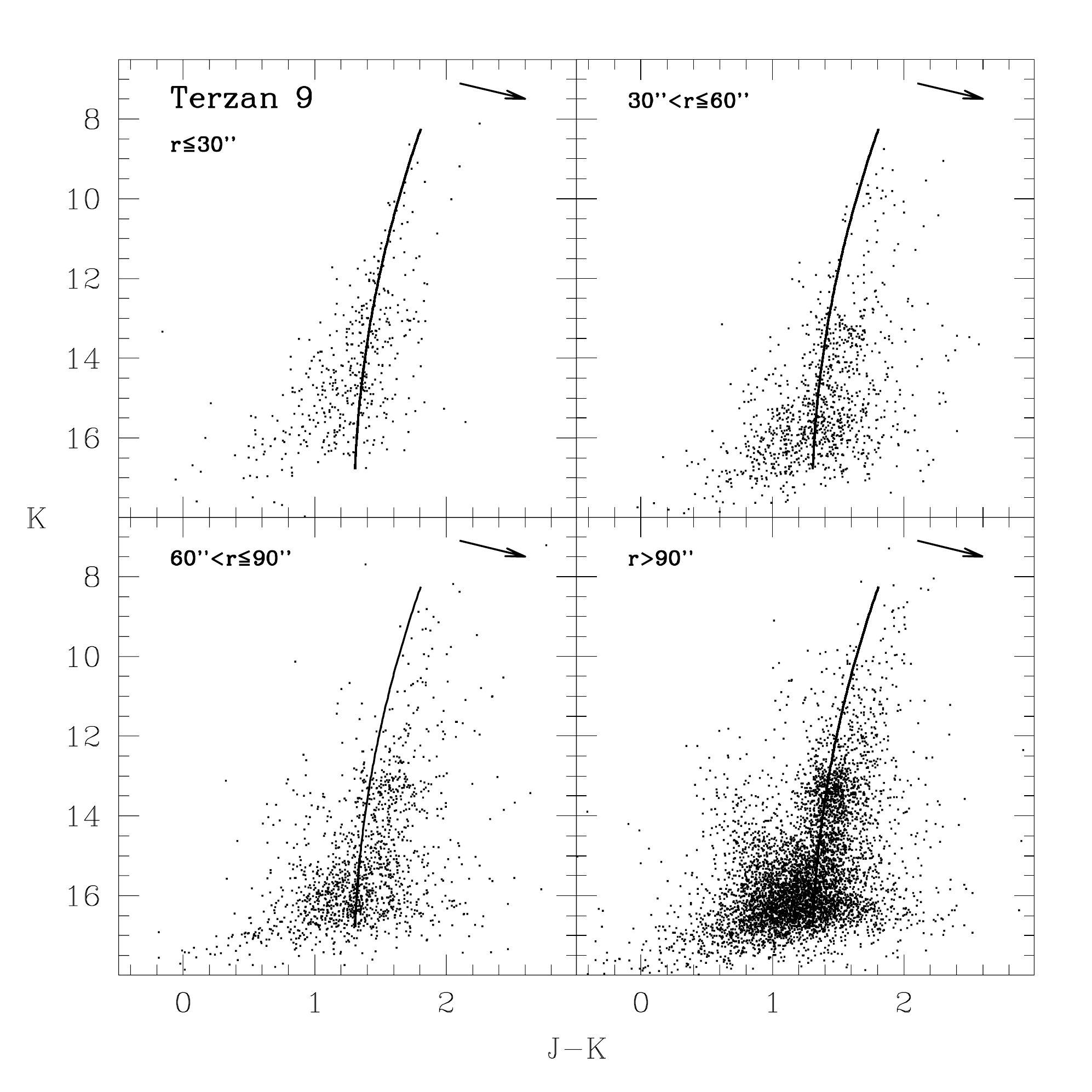}
\caption{[K, J--K] CMD of Terzan~9 at different distances (r) from the cluster center. The arrow represents the reddening vector.}
\label{ter9}
\end{figure}

Terzan~9 is another example of a heavily reddened cluster [E(B\--V)=1.87, \citet{har}] in the inner Bulge, whose metallicity estimates in the literature span a large range from [Fe/H]=--0.38~dex \citep{zin85} to [Fe/H]=--1.01~dex \citep{bica98}
 The only photometry available so far is the optical study by \citet{ter96453} which provides measurements of the cluster reddening [E(B\--V)=1.95] and distance [(m\--M)$_0$=13.45] by comparing the cluster CMD in the [V, V\--I] plane with that of M~30. 
The dominant features of the IR CMD (see Fig.~\ref{cmds}) are the red clumpy HB at K$\sim$13.5, the RGB which appears to be quite scattered, and the main sequence of disk field at 12.5$<$K$<$16 and 0.8$<$(J\--K)$<$1.2. However, the radial CMDs (see Fig.~\ref{ter9}) clearly show that the innermost cluster region (r$\leq$30$\arcsec$) is characterized by a narrow RGB and a blue HB. In particular, a
blue HB was also detected in the optical photometry by \citet{ter96453}.
The HB red clump starts to be visible at distances larger than 30$\arcsec$, when also the RGB gets redder and
scattered. As in the case of Ter~1 and Ter~4, at distances larger than r$>60\arcsec$ the bulge field population dominates the IR CMD, responsible for the spread in the RGB and for the presence of the HB red clump.

In order to derive the cluster properties we used the FVO06 empirical method, assuming the bulgelike scenario. We measured the RGB slope considering only those stars lying in the innermost cluster region (RGB$_{slope}$=--0.079), and we estimated the observed (K= 8.131) RGB tip using the brightest star in the catalog lying along the RGB ridgeline. We used these values as input parameters for the computational routine, and we obtained the following reddening, distance and metallicity estimates: E(B\--V)=1.79, (m\--M)$_0$=13.73, [Fe/H]=--1.21~dex and [M/H]=--1.01~dex.
The reddening estimate is in fair agreement with that derived from the optical study, while the distance modulus difference of $\approx$0.25~mag is more likely due to a different assumption of the distance scale, from \citet{frf99} here, and from \citet{barb98} in \citet{ter96453}.
Our photometric metallicity value is slightly lower than the literature values. However only high resolution spectroscopy could firmly establish the cluster metallicity since the previous studies [e.g. \citet{bica98}] based on integrated IR spectroscopy could have been affected by the high level of bulge field contamination which would then result in a spurious higher value for the cluster metallicity.

\subsection{HP~1}

\begin{figure} 
\includegraphics[width=8cm]{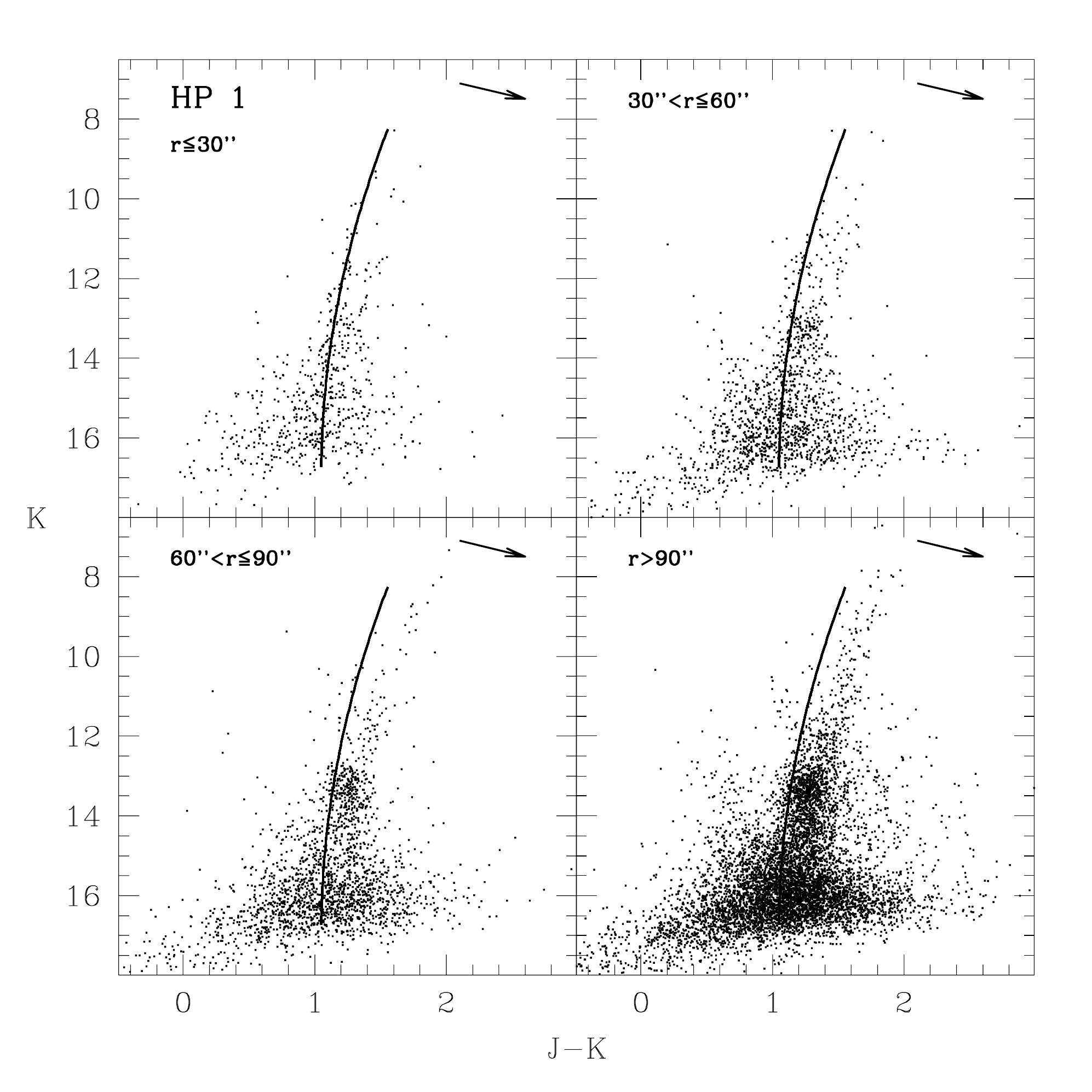}
\caption{[K, J--K] CMD of HP~1 at different distances (r) from the cluster center. The arrow represents the reddening vector.}
\label{hp1}
\end{figure}

HP~1 has been subject of both optical \citep{hp1_B} and IR \citep{hp1lil1djo1} ground based photometry. The optical CMD of HP~1 shows an extended blue HB, suggesting a low\--intermediate metallicity, which has been later on confirmed by \citet{hp1_A} who performed high resolution optical spectroscopy. This latter study found the cluster to be metal\--intermediate ([Fe/H]=--1.00~dex) and
$\alpha$--elements enhanced ([$\alpha$/Fe]=+0.3).
From the cluster CMD, \citet{hp1_B} measured the cluster reddening E(B\--V)=1.19 and distance (m--M)$_0$=14.15, being both values quite different from those derived by \citet{hp1lil1djo1} [namely, E(B\--V)=0.74 and (m\--M)$_0$=14.70]. 

As for the other clusters, the IR CMD of HP~1 shown in Fig.~\ref{cmds} is dominated by a red clumpy HB
and a scattered RGB. The analysis of the radial CMDs (see Fig.~\ref{hp1})  reveals once again  that these two features are due to the bulk of the bulge field population. In fact, the RGB in the innermost cluster region (r$\leq$30$\arcsec$) is quite blue, narrow and steep, and the HB is barely visible as a blue vertical sequence at 14$<$K$<$16 and 0$<$(J--K)$<$1. With increasing distance from the cluster center, the contamination by bulge field stars makes the RGB redder and more bent, and populates the red HB clump. Measuring  the RGB slope in the innermost cluster region (RGB$_{slope}$=--0.083) and assuming for the RGB tip (K$^{tip}$=8.287) the magnitude of the brightest star along the ridgeline, the computational routine gives E(B\--V)=1.18 for the reddening, (m\--M)$_0$=14.17 for the distance, [Fe/H]=--1.12~dex and [M/H]=--0.91~dex for the metallicity.

\subsection{Liller~1}
Among the observed clusters, Lil~1 is the most metal\--rich and also the most reddened. 
Given its high level of extinction (A$_V>$9), this cluster has been mostly investigated in IR.
Based on high\--resolution IR spectroscopy, \citet{lil1_B} found [Fe/H]=--0.30~dex and [$\alpha$/Fe]=+0.3~dex. The only optical photometric study is that of \citet{lil1_C} who measured the cluster reddening E(B\--V)=3.05 and  distance (m--M)$_0$=14.52. Similar values [E(B\--V)=3.0, (m--M)$_0$=14.68] have been measured also by \citet{lil1_D} who sampled the cluster RGB by using IR photometry. Later on, \citet{hp1lil1djo1} provided deeper JHK photometry, down to K$\sim$18.5 from which they estimated a reddening E(B\--V)=3.13 and a metallicity close to that of NGC~6528, the most metal\--rich Bulge cluster.
Recently, by using NICMOS on board of HST, \citet{lil1ter4_A} obtained deep photometry which provided an estimate of the cluster age, via the main sequence turn\--off measurement. The author found Liller~1 to be coeval to 47~Tuc.

Our IR CMD (see Fig.~\ref{cmds}) of Liller~1 samples the extension of the RGB, from the tip down to the HB. The morphology of the RGB (really bent) and of the HB (clumpy) suggest a high metallicity. When we compare the cluster CMDs and LFs with those of 47~Tuc, we derived E(B\--V)=3.09 and (m\--M)$_0$=14.48. Both estimates nicely agree with the corresponding values found by previous studies.
The set of IR photometric indices measured along the RGB ridgelines gives a cluster metallicity of [Fe/H]=--0.36~dex and [M/H]=--0.14, which are fully consistent with the estimates obtained by \citet{lil1_B}.

\subsection{Djorgovski~1}

\begin{figure} 
\includegraphics[width=8cm]{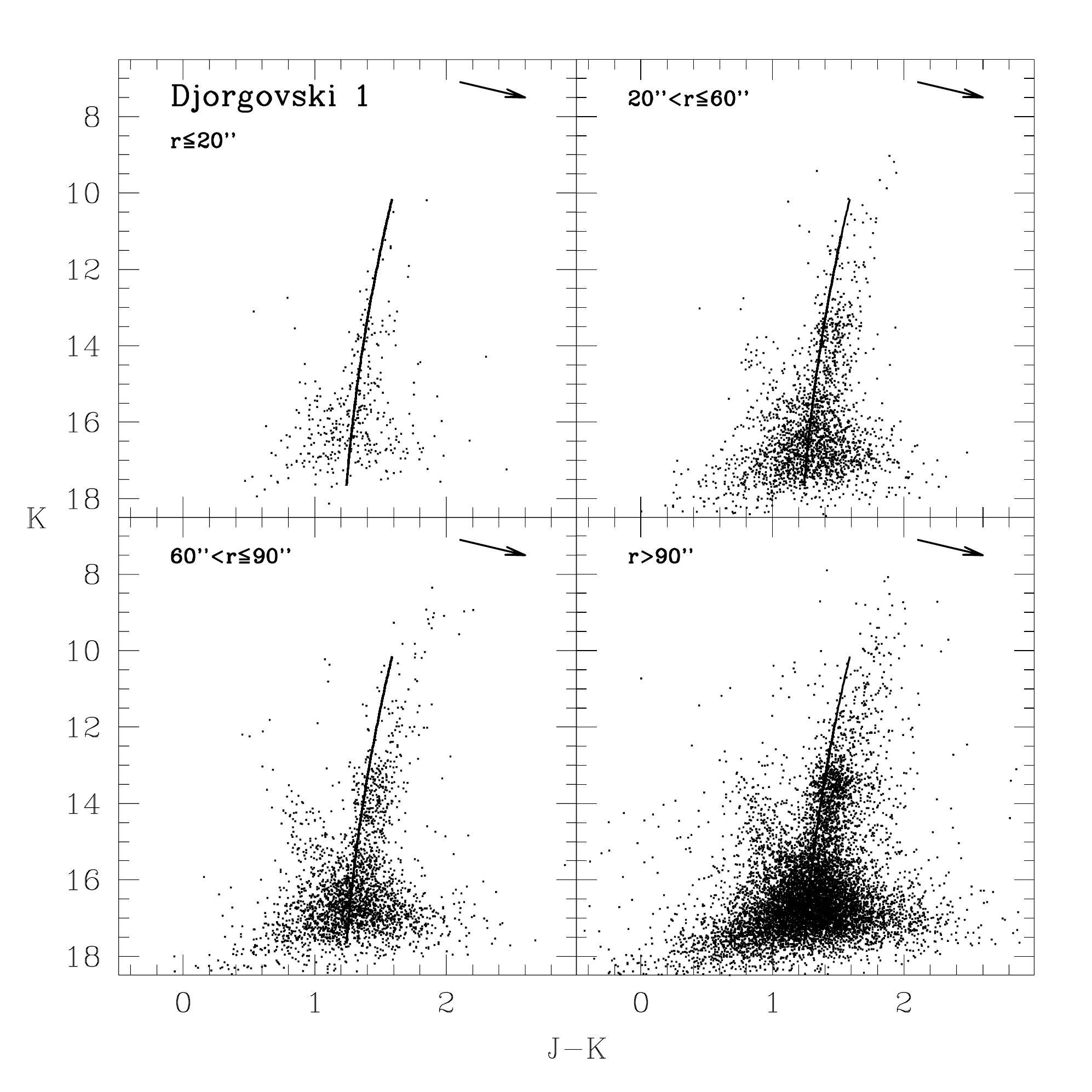}
\caption{[K, J--K] CMD of Djo~1 at different distances (r) from the cluster center. The arrow represents the reddening vector.}
\label{djo1}
\end{figure}

Djorgovski~1 is a loose [c=1.50, \citet{tra93}] cluster discovered in 1987 by Djorgovski during an observing campaign aimed at identify the optical counterparts to IRAS sources. It has been observed in optical by \citet{djo1_A} and in IR by \citet{hp1lil1djo1}. However, the two studies reached very different conclusions regarding the main cluster properties. In fact, by using [V, V\--I] CMD \citet{djo1_A} derived E(B\--V)=1.71, (m\--M)$_0$=14.72 and [Fe/H]=--0.40~dex, for the cluster reddening, distance and metallicity, respectively.
From the IR CMD, \citet{hp1lil1djo1} measured a slighly lower reddening [E(B\--V)=1.44] and a considerably longer distance [(m\--M)$_0$=15.40]. Moreover, by comparing the CMD of Djo~1 with
that of M~92, the authors concluded that Djo~1 must be extremely metal\--poor. Unfortunately no high resolution spectroscopy is available for this cluster.

Our IR CMD (see Fig.~\ref{cmds}) shows the presence of a red HB clump and a large spread in the RGB. A careful inspection of the innermost ($r\leq20$") region (see Fig.~\ref{djo1}) shows a quite narrow blue RGB, without any evidence of red clump HB stars. The HB red clump starts to be populated with increasing distances from the cluster center, when also the RGB becomes redder and more spread because of the field contamination.
Reddening, distance and metallicity estimates of [E(B\--V)=1.58, (m\--M)$_0$=15.65, [Fe/H]=--1.51~dex and [M/H]=--1.31~dex] were obtained using the empirical method of FVO06, with the following input parameters: RGB$_{slope}$=--0.068 and K$^{tip}$=10.152. Our finding are consistent within the errors with the IR study of \citet{hp1lil1djo1}.

\subsection{Djorgovski~2}
Djorgovski~2, also known as ESO~456--SC~38, is a rather loose cluster with no evidence of post\--core collapse morphology \citep{tra93}. The only available photometry is that provided by \citet{djo2_A} 
which lead to the following estimates for the cluster reddening and distance, respectively: E(B\--V)=0.98, (m\--M)$_0$=13.70. The authors claimed that Djo~2 should have approximately the same metallicity as 47~Tuc, thus somewhat lower that the value measured by \citet{bica98} ([Fe/H]=--0.40~dex), based on integrated IR spectroscopy.

The CMD shown in Fig.~\ref{cmds} represents the first IR photometry of Djo~2 ever obtained. The CMD of the cluster doesn't shows a large field contamination as in the case of the majority of the other sampled clusters. The RGB is quite narrow and bent, and together with the HB morphology suggests a high metallicity. From the cluster CMDs and LFs we estimated a reddening E(B\--V)=0.94 and distance (m\--M)$_0$=14.23. The photometric indices measured along the RGB ridgeline in the [M$_K$, (J\--K)$_0$] and [M$_H$, (J\--H)$_0$] absolute planes (see Tables~\ref{jkRGB} and \ref{jhRGB}), yielded the following metallicity estimates: [Fe/H]=--0.65~dex and [M/H]=--0.45~dex.

\subsection{NGC~6540}

\begin{figure} 
\includegraphics[width=8cm]{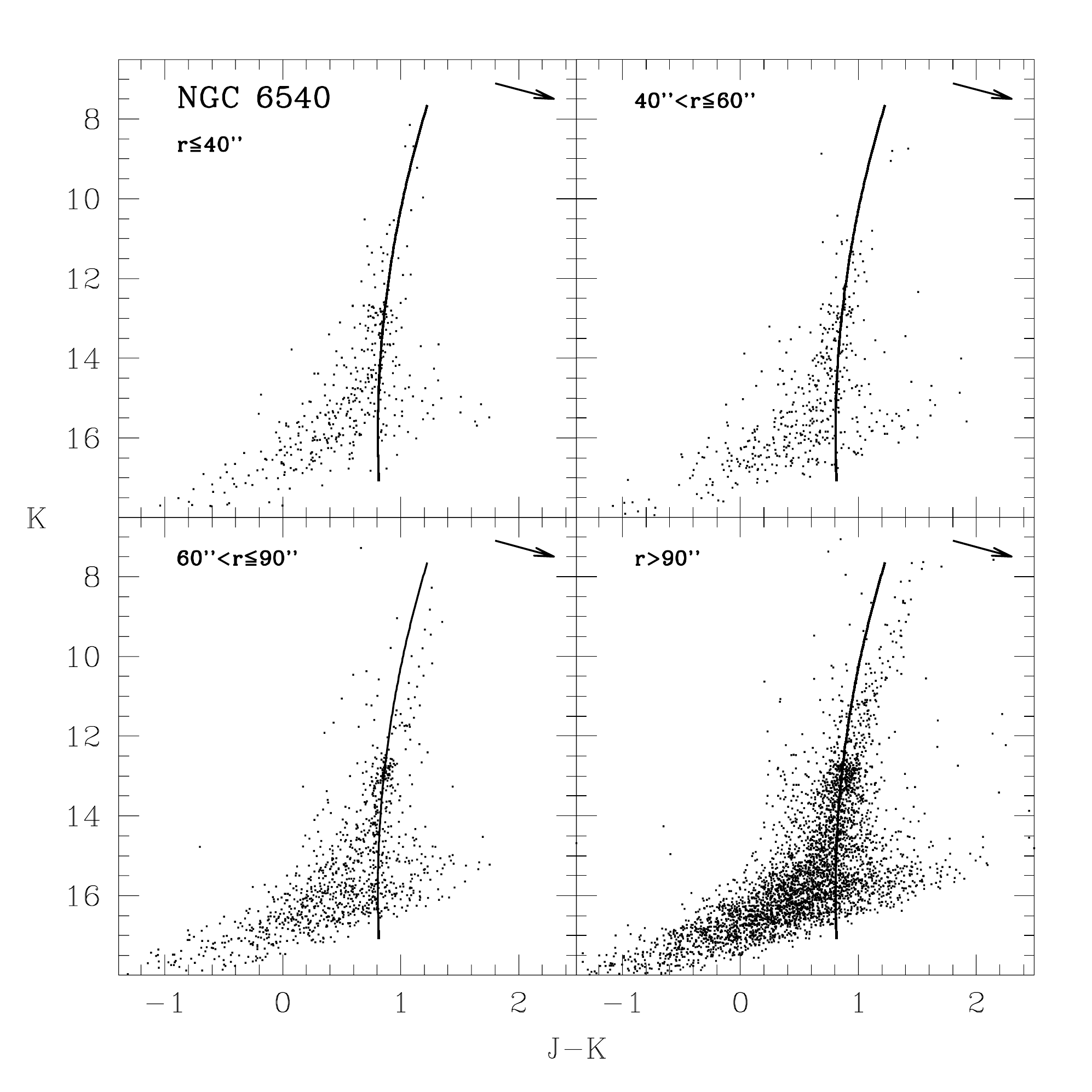}
\caption{[K, J--K] CMD of NGC~6540 at different distances (r) from the cluster center. The arrow represents the reddening vector.}
\label{6540}
\end{figure}
NGC~6540 has been classified as an open cluster until  \citet{6540_A} provided the first optical CMD, showing that they were dealing with a GC. From their photometry, the authors derived a reddening E(B\--V)=0.60 and a distance modulus (m\--M)$_0$=12.70. From optical integrated spectroscopy, they estimate [Fe/H]=--1.0.

Fig.~\ref{cmds} presents the first IR CMD of the cluster, whose most prominent features are the presence of a red HB and a large scattered RGB [(J\--K)$\sim$0.5]. However, the radial CMDs (see Fig.~\ref{6540}) show that the red HB doesn't belong to the cluster, but to the bulge field population. Indeed, the CMD of the cluster (at distance r$\leq$60$\arcsec$) is characterized by a steep poorly populated RGB and by a blue HB at 13$<$K$<$16 and -0.2$<$(J\--K)$<$0.7, also present in the optical CMD of \citet{6540_A}. At larger distances from the cluster center (r$>$60$\arcsec$) the field population dominates the CMD, being responsible for the morphology of the HB and RGB, the latter becoming redder, more bent and scattered because of metallicity and depth effects.
The RGB ridgeline shown in Fig.~\ref{cmds} and \ref{6540} has been derived by using only stars within the innermost cluster region, and the cluster properties were estimated using the empirical routine of FVO06. From the observed CMD we measured RGB$_{slope}$=--0.068 and K$^{tip}$=7.584. Adopting these two values we found E(B\--V)=0.66, (m\--M)$_0$=13.57, [Fe/H]=--1.29~dex and [M/H]=--1.10~dex for the cluster reddening, distance, and metallicity.

\subsection{NGC~6544}
NGC~6544 is a very poorly studied cluster of the inner Bulge. The only photometry available is that by \citet{6544_A} who provided a CMD with 160 stars measured on photographic plates. The published CMD is poorly populated and it shows a blue HB, suggesting a medium metallicity.
The study of \citet{6544_A} provived the first estimate of the cluster reddening E(B\--V)=0.70 and distance d$_{\odot}$=2.8~kpc. 

Our IR CMD (see Fig.~\ref{cmds}) samples the cluster RGB from the tip down to $\sim$2~mag below the main sequence turn\--off. It also shows that even if the field contamination is high, it is not an issue being significantly redder than the cluster sequences. In fact, the cluster is characterized by a narrow, steep and blue [0.4$,$(J\--K)$<$0.7] RGB , and by a blue HB visible as the vertical sequence at 11.5$<$K$<$14 and 0.3$<$(J\--K)$<$0.6. The field population is instead responsible for the red RGB sequence and for the red HB, being more metal\--rich and more reddened. Because the cluster HB morphology does not allow a reliable determination of the HB level, we measured the cluster properties making use of the empirical method and with the following input parameters: RGB$_{slope}$=--0.073 and K$^{tip}$=6.242. We obtained E(B\--V)=0.78, (m\--M)$_0$=12.23, [Fe/H]=--1.19~dex and [M/H]=--1.00~dex, for the cluster reddening, distance and metallicity, which are consistent with the previous optical study.
From the difference in color between the two RGBs in Fig.~\ref{cmds}, the field population should be affected by a higher reddening E(B\--V)=1.12 and more distant than the cluster.

\subsection{NGC~6522}

\begin{figure} 
\includegraphics[width=8cm]{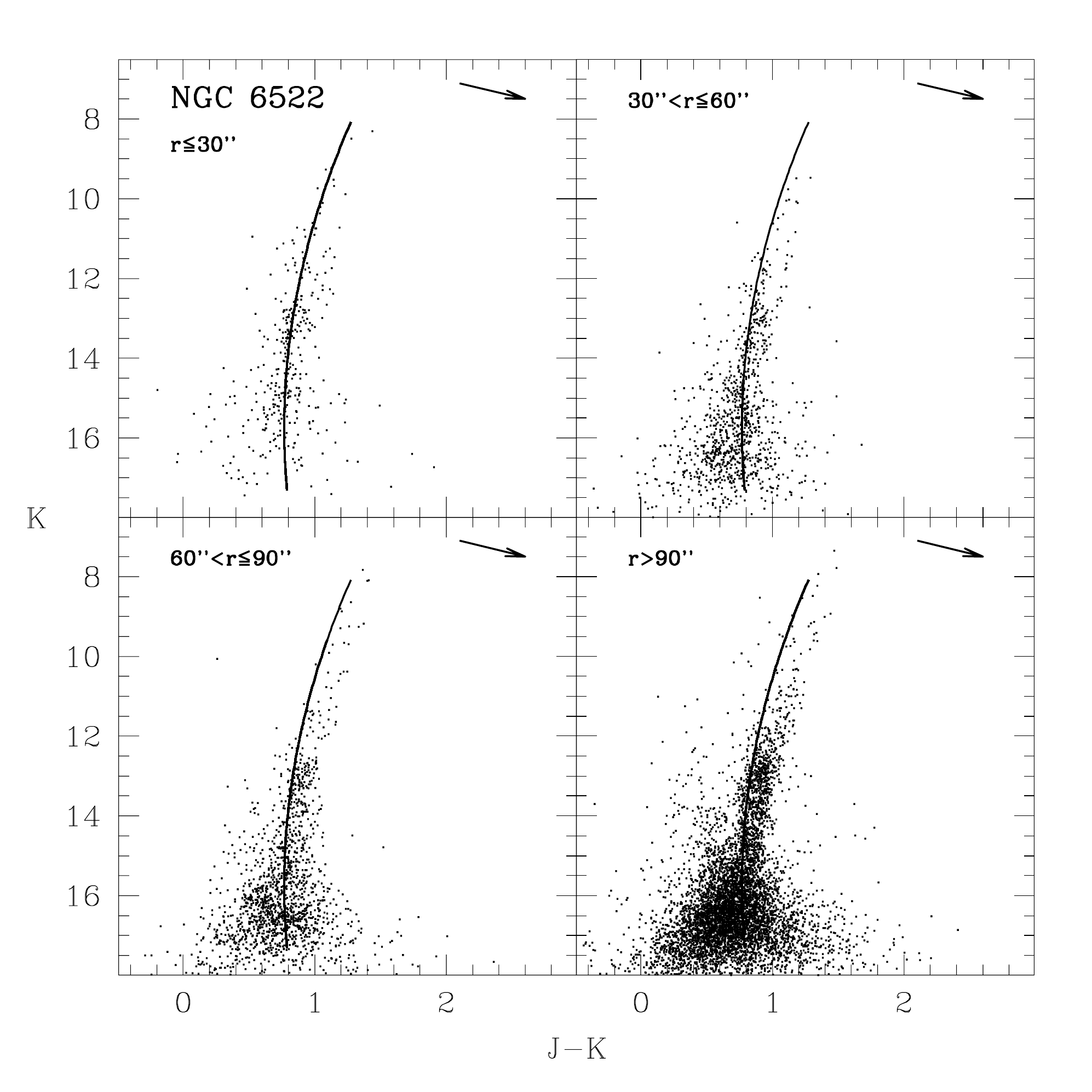}
\caption{[K, J--K] CMD of NGC~6522 at different distances (r) from the cluster center. The arrow represents the reddening vector.}
\label{6522}
\end{figure}

NGC~6522 is located close to the low extinction region known as Baade's Window, which is the most studied region of the Galactic bulge. Because of its location and the presence of a large sample of blue HB stars, this cluster has been subject of many studies \citep{6522_A,6522_B,6522_C,6522_D,6522_E, dav00,  dav01} aimed at deriving the main cluster properties. Integrated light studies provide metallicities in the range --1.44$<$[Fe/H]$<$--1.05~dex and reddening 0.45$<$E(B\--V)$<$0.6 \citep{bipas83,zw84}. 

Our IR CMD (see Fig.~\ref{cmds}) reveals the presence of a quite\--scattered RGB, which cannot be explaned in terms of photometric errors. As done previously, in order to understand the nature of the observed scatter in Fig.~\ref{6522} we compared the radial CMDs in four annuli at different distances from the cluster center. In the innermost cluster region (r$<30\arcsec$) a blue HB at 13$<$K$<$14.5 becomes visible and the RGB appears quite narrow, and progressively spreads out with increasing distance from the center. A field component redder than the cluster RGB mean loci starts to be important at r$>$60$\arcsec$.
By using the empirical method of FVO06 with the following input parameters, RGB$_{slope}$=--0.067 and K$^{tip}$=8.495, we derived E(B\--V)=0.66, (m\--M)$_0$=14.34, [Fe/H]=--1.52~dex and [M/H]=--1.33~dex Our metallicity and reddening estimates are consistent with the values found in literature, while the distance nicely agree with that published by \citet[][d$_{\odot}$=7.3~kpc]{6522_B} and within the values published by \citet[][(m\--M)$_0$= 14.1 \-- 14.6, for different calibrations]{dav00}.

\subsection{NGC~6453}
As NGC~6540, NGC~6453 is also a poorly studied GC.  \citet{ter96453} derived the reddening E(B\--V)=0.70 and distance d$_{\odot}$=7.9~kpc from [V, V\--I] CMD. While the reddening estimate is in nice agreement with the value quoted by \citet{zw84,bica86} (0.60, 0.61, respectively), the distance derived by \citet{ter96453} is considerable shorter than that found in the \citet{har} compilation (d$_{\odot}$=10.5~kpc). Later on, by using [K, J\--K] CMD \citet{dav00} obtained E(B\--V)=0.63, [Fe/H]=--1.9 and absolute distance modulus (m\--M)$_0$=14.9 \-- 15.1, for different calibrations. 

\begin{figure} 
\includegraphics[width=8cm]{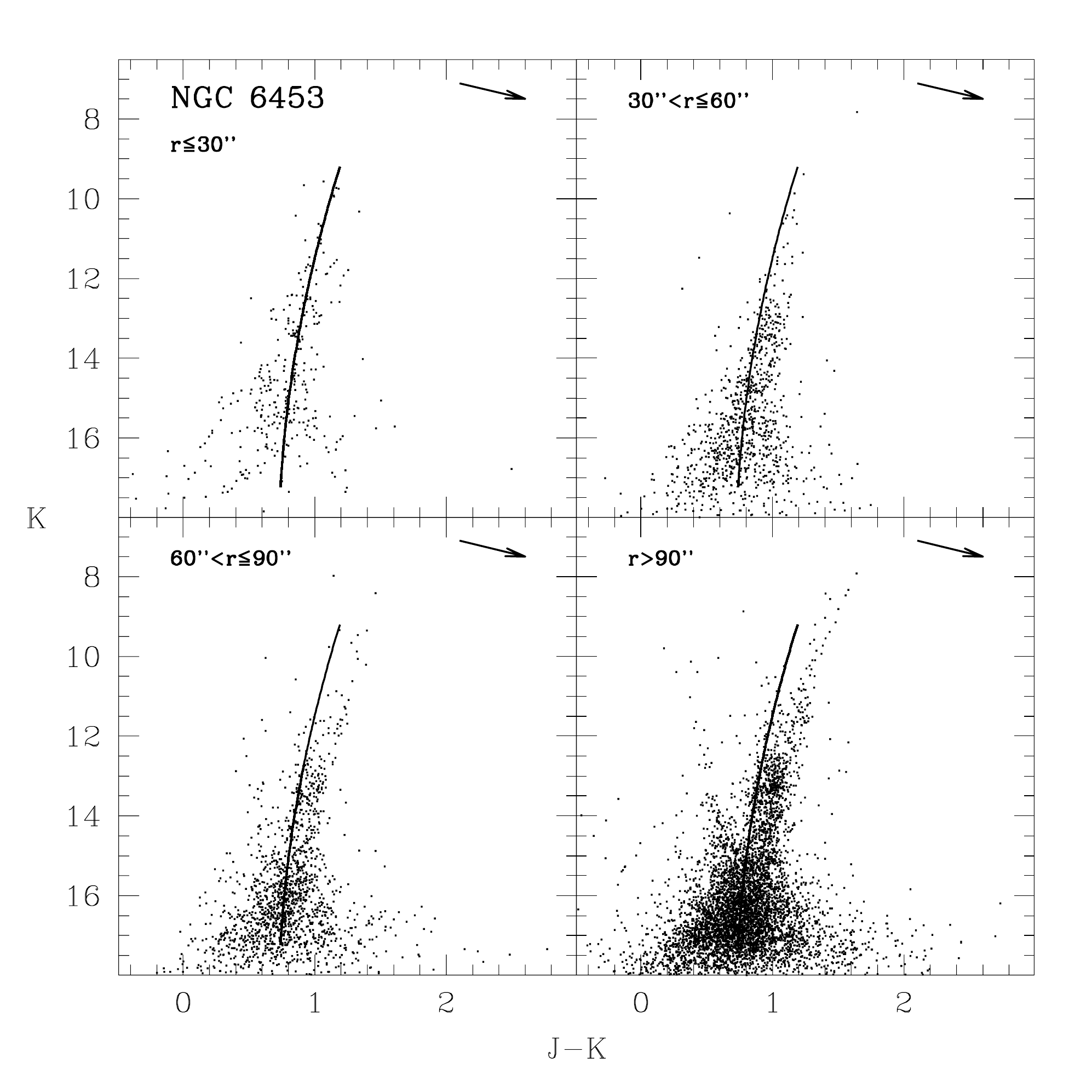}
\caption{[K, J--K] CMD of NGC~6453 at different distances (r) from the cluster center. The arrow represents the reddening vector.}
\label{6453}
\end{figure}

Our IR CMD of NGC~6453 (see Fig.~\ref{cmds}) is very similar to that of NGC~6522, with a scattered RGB which seems to be split in two distinct sequences and a clumpy and elongated HB.
However, when considering only stars in the innermost center region (see Fig.~\ref{6453}) the CMD is characterized by a blue steep RGB with no evidence of a red clump HB, thus suggesting a low\--intermediate metallicity.  The red HB starts {\bf to }appear at increasing distances from the cluster center (r$>30\arcsec$) together with the red RGB component.
The RGB fiducial ridgeline was derived using stars in the innermost cluster region where the field contamination is low. The measured reddening, distance and metallicity estimates [E(B\--V)=0.69, (m\--M)$_0$=15.15, [Fe/H]=--1.57~dex and [M/H]=--1.38~dex] were obtained running the FVO06's routine with RGB$_{slope}$=--0.075 and K$^{tip}$=9.337 as input parameters. Our distance estimate (d$_{\odot}$=10.7~kpc) is consistent within the error with those of \citet{har} and \citet{dav00}, while the metallicity nicely agrees to what \citet{bipas83} found from visible and IR low resolution spectroscopy ([Fe/H]=--1.40~dex). Also the reddening derived in this work is fully consistent with the values found in the previous studies quoted above.

\begin{figure} 
\includegraphics[width=8cm]{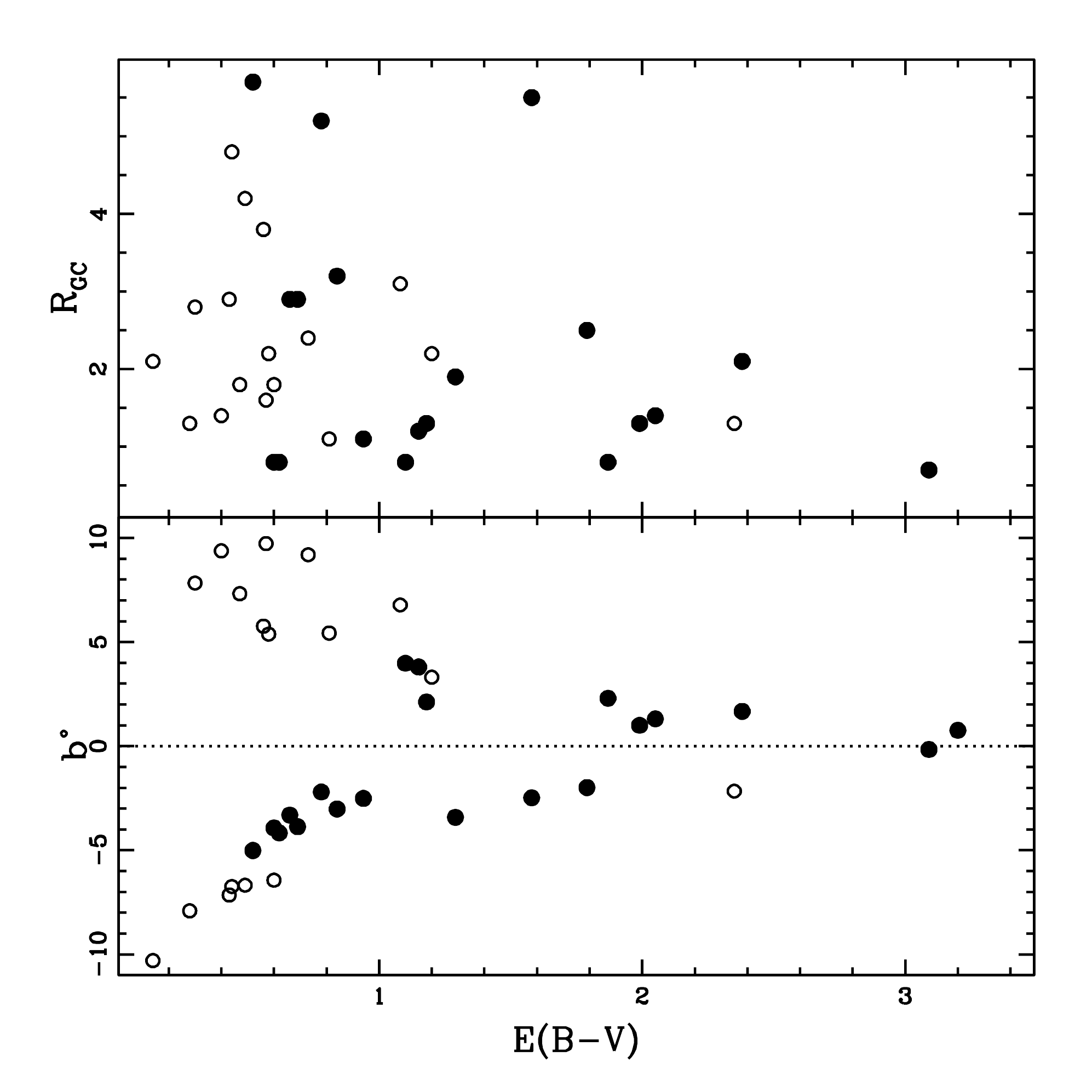}
\caption{Distribution of the clusters extinction as a function of the distance to the Galactic center and of the Galactic latitude ({\it upper} and {\it lower panels}, respectively) as derived in this work and in {\it Paper~I}. Filled and empty circles mark the inner and outer Bulge clusters, respectively.}
\label{ebv}
\end{figure}

\section{Discussion and Conclusion}
In this section we briefly discuss the main properties of the global sample of Bulge GCs collected
in our survey, counting 24 clusters presented in {\it Paper~I} and 12 discussed in this work.
It is worth mentioning that Tables~\ref{param} and \ref{bol} of this work and Tables 1 and 2 of {\it Paper~I} list the largest homogeneous catalog of Bulge GCs obtained so far, whose properties have been derived in a fully self\--consistent way. Moreover, the entire sample represents ~80\% of the clusters population in the Bulge direction.
\begin{figure} 
\includegraphics[width=8cm]{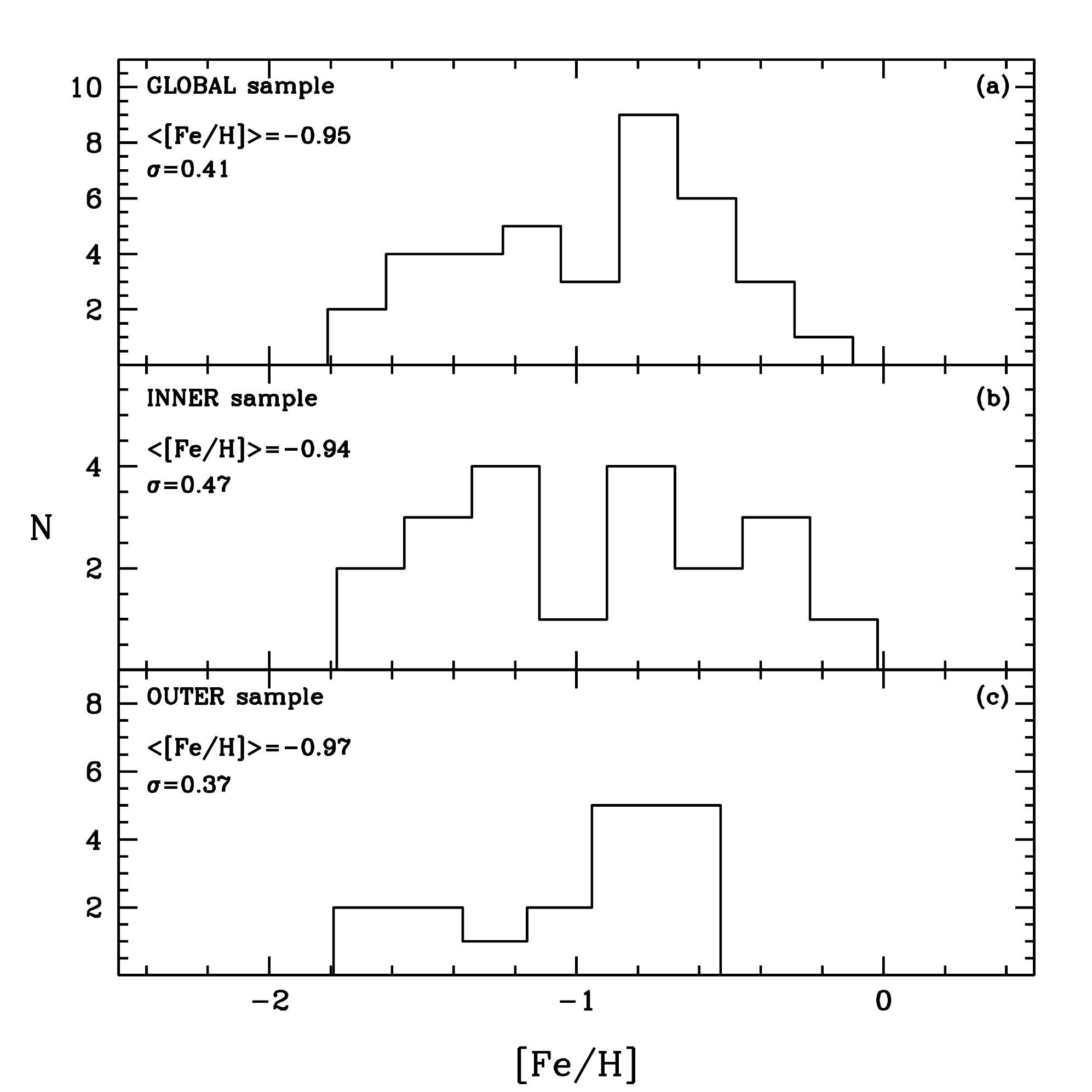}
\caption{Photometric Metallicity Distribution as derived in this work and in {\it Paper~I} for the
global clusters sample [{\it panel (a)}], and for the inner and outer clusters [{\it panel (b)} and
 {\it panel (c)}], respectively.}
\label{md}
\end{figure}

In Figure~\ref{ebv} the distribution of the extinction for the global cluster sample (this work and {\it Paper~I}) as a function of the distance to the Galactic center and of the Galactic latitude is shown. As expected, the presence in the Bulge of highly variable and patchy extinction is also traced by the clusters reddening distribution, which shows a progressively drops of the reddening values from the inner towards the outer Bulge regions (see lower panel of Fig.~\ref{ebv}), and a larger scatter in the inner 2~Kpc.

\begin{figure} 
\includegraphics[width=8cm]{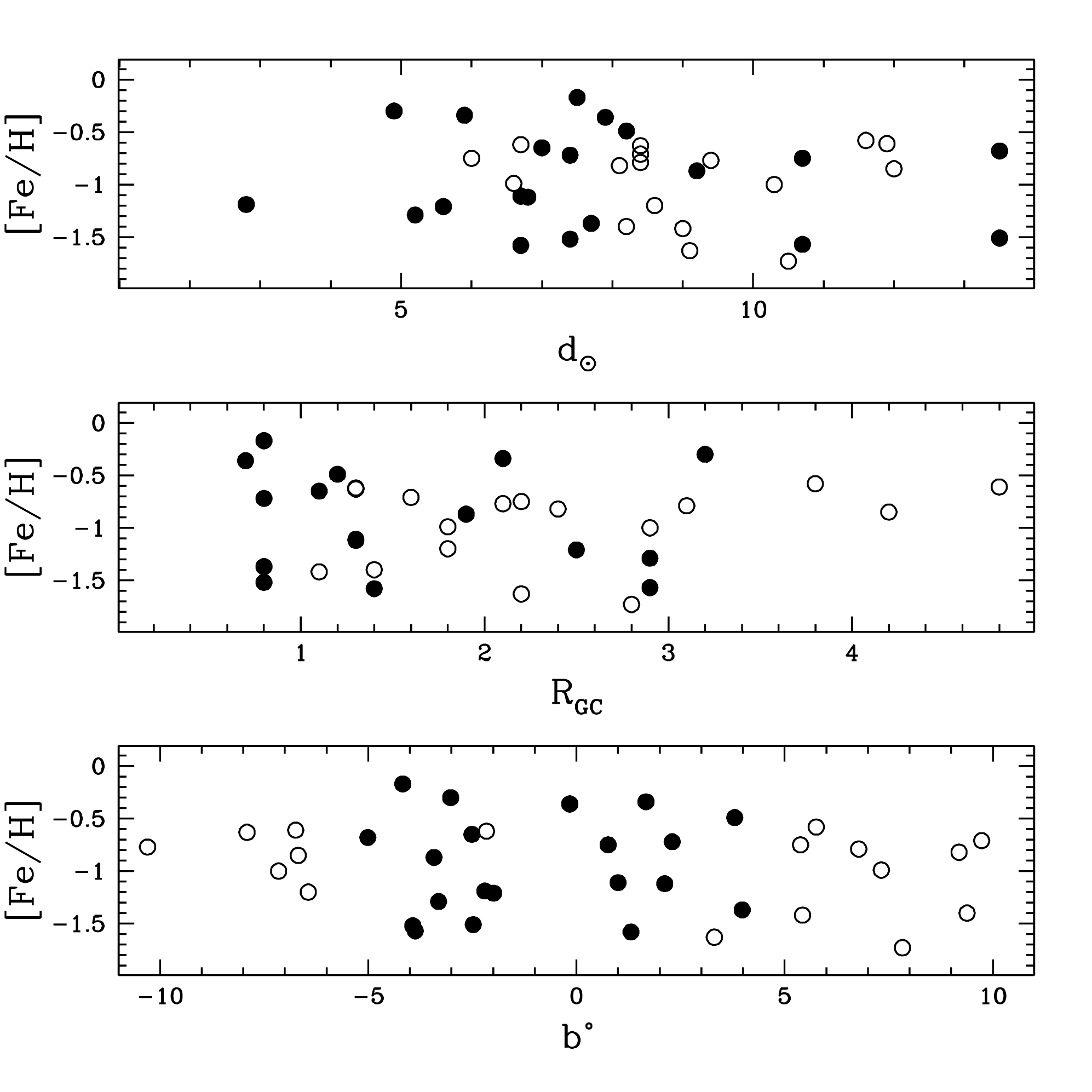}
\caption{MD of the global cluster sample as a function of the distance to the Sun ({\it upper panel}); to
the Galactic center ({\it middle panel}); and the Galactic latitude ({\it lower panel}). Filled and empty symbols represent the inner and outer Bulge clusters, respectively.}
\label{multi}
\end{figure}

Figure ~\ref{md} ({\it panel a)}) shows the Metallicity Distribution (MD) of the global cluster sample as derived from our photometric studies. The most striking feature of the MD is the presence of two distinct peaks at [Fe/H]$\sim$--1.4~dex and [Fe/H]$\sim$--0.7~dex. When the MD of the inner ($|l|{\leq}10^{\circ}$ and  $|b|{\leq}5^{\circ}$) and outer ($|l|>10^{\circ}$ and  $|b|>5^{\circ}$) Bulge clusters are considered separately (see Fig.~\ref{md}, {\it panels (b)} and {\it (c)}), one finds that the metal\--poor clusters are equally distributed within the Bulge. Conversely, a remarkable difference is present between the inner and outer metal\--rich population. While in the outer Bulge no cluster with [Fe/H]$\geq$-0.5~dex is found, the inner region is populated by clusters with metallicity up to solar value.

Since recent studies based on high resolution optical and IR spectroscopy of Bulge field giants found
indications of a mild radial metallicity gradient between b$=-4^{\circ}$ and  b$=-12^{\circ}$ \citep[see][]{zoc08} and a flattening between $(l,b)=(1, -4)$ and $(l,b)=)(0, -1)$ \citep[see][]{rov07}, we used our
catalog of cluster properties to check the presence of possible correlations between the clusters metallicity and the radial distance as shown in Figure~\ref{multi}. We found no evidences for a radial metallicity gradient. The clusters span a large metallicity range at any distance to the Sun (see upper panel of Fig.~\ref{multi}), to the Galactic center (see middle panel of Fig.~\ref{multi}), and at any Galactic latitude (see  lower panel of Fig.~\ref{multi}).

\begin{table*}
\begin{center}
\caption{RGB location in colour (column 4-7), in magnitude (column 8) in the K, J\--K plane and the RGB slope for the cluster sample}
\label{jkRGB}
\begin{tabular}{lcccccccc}
\hline \hline
Name& $(m\--M)_0$& $E(B\--V)$ & $(J\--K)_0^{-5.5}$ &  $(J\--K)_0^{-5}$ &  $(J\--K)_0^{-4}$ &  $(J\--K)_0^{-3}$ & $M_K^{(J\--K)_0=0.7}$ &$RGB_{slope}$ \\
\hline
Ter1 &	14.13 &   1.99 &    0.876 &   0.820 &	0.722 &   0.645 &   -3.752 &  -0.079 \\
Ter2 &	14.35 &   1.87 &    0.971 &   0.913 &	0.808 &   0.720 &   -2.737 &  -0.102 \\
Lil1 &	14.48 &   3.09 &    1.099 &   1.036 &	0.922 &   0.819 &   -1.677 &  -0.111 \\
Djo2 &	14.23 &   0.94 &    1.036 &   0.980 &	0.879 &   0.791 &   -1.772 &  -0.101 \\
\hline
\end{tabular}
\end{center}
\end{table*}

\begin{table*}
\begin{center}
\caption{RGB location in colour (column 4-7), in magnitude (column 8) in the H, J\--H plane and the RGB slope for the cluster sample}
\label{jhRGB}
\begin{tabular}{lcccccccc}
\hline \hline
Name& $(m\--M)_0$& $E(B\--V)$ & $(J\--H)_0^{-5.5}$ &  $(J\--H)_0^{-5}$ &  $(J\--H)_0^{-4}$ &  $(J\--H)_0^{-3}$ & $M_H^{(J\--H)_0=0.7}$ &$RGB_{slope}$ \\
\hline
Ter1 &	14.13 &   1.99 &      0.763 &	0.726 &   0.662 &   0.609 &   -4.611 &  -0.053\\
Ter2 &	14.35 &   1.87 &      0.794 &	0.755 &   0.683 &   0.621 &   -4.247 &  -0.078\\
Lil1 &	14.48 &   3.09 &      0.846 &	0.810 &   0.744 &   0.686 &   -3.265 &  -0.086\\
Djo2 &	14.23 &   0.94 &      0.806 &	0.756 &   0.667 &   0.593 &   -4.384 &  -0.070\\
\hline
\end{tabular}
\end{center}
\end{table*}

\begin{table*}
\begin{center}
\caption{Coordinates, Distance, Reddening and Metallicity for the sample of Bulge clusters}
\label{param}
\begin{tabular}{lccccccccccc}
\hline \hline
Name& l & b & d$_{\odot}$ & R$_{GC}$ & X & Y & Z &(m-M)$_0$ &E(B-V) &[Fe/H] &[M/H] \\
  & (deg) & (deg) & (Kpc) &(Kpc) & (Kpc) &(Kpc) &(Kpc) &(mag)&(mag) &(dex) &(dex) \\
(1)&(2)&(3)&(4)&(5)&(6)&(7)&(8)&(9)&(10)&(11)&(12)\\
\hline
 Ter~1 &  -2.43   &  1.00   &  6.6  &  1.5  & 6.5  & -0.3  &  0.1   &  14.13  &  1.99  &   -1.11 &  -0.91  \\
 Ter~2 &  -3.68   &  2.30   &  7.4  &  0.8  & 7.3  & -0.5  &  0.3   &  14.35  &  1.87  &   -0.72  &  -0.53 \\
 Ter~4$^*$ &  -3.98   &  1.31   &  6.7  &  1.4  & 6.6  & -0.5  &  0.2   &  14.13  &  2.05  &   -1.58  &  -1.37 \\
 Ter~9$^*$ &   3.60   & -1.99   &  5.6  &  2.5  & 5.5  &  0.3  & -0.2   &  13.73  &  1.79  &   -1.21  &  -1.01 \\
  HP~1$^*$ &  -2.58   &  2.12   &  6.8  &  1.3  & 6.8  & -0.3  &  0.3   &  14.17  &  1.18  &   -1.12  &  -0.91 \\
  Lil~1  &  -5.16   & -0.16   &  7.9  &  0.7  & 7.8  & -0.7  & -0.0   &  14.48  &  3.09  &   -0.36  &  -0.14\\	
  Djo~1$^*$ &  -3.33   & -2.48   & 13.5  &  5.5   &13.4  & -0.8  & -0.6   &  15.65  &  1.58  &   -1.51  &  -1.31\\ 
  Djo~2 &   2.76   & -2.51   &  7.0  &  1.1 & 7.0  &  0.3  & -0.3   &  14.23  &  0.94  &   -0.65  &  -0.45 \\
 NGC~6540$^*$ &   3.29   & -3.31   &  5.2  &  2.9  & 5.2  &  0.3  & -0.3   &  13.57  &  0.66  &   -1.29  &  -1.10  \\
 NGC~6544$^*$ &   5.84   & -2.20   &  2.8  &  5.2  & 2.7  &  0.3  & -0.1   &  12.23  &  0.78  &   -1.19  &  -1.00 \\
 NGC~6522$^*$ &   1.02   & -3.93   &  7.4  &  0.8  & 7.3  &  0.1  & -0.5   &  14.34  &  0.66  &   -1.52  &  -1.33 \\
 NGC~6453$^*$ &  -4.28   & -3.87   & 10.7  &  2.9  &10.6  & -0.8  & -0.7   &  15.15  &  0.69  &   -1.57  &  -1.38\\ 
&&&&&&&&&&&\\
\hline
\multicolumn{12}{l}{Notes \-- l and b are from \citet{har} catalog. Asterisk signify clusters for which the metallicity estimates were}\\
\multicolumn{12}{l}{obtained using the FVO06 empirical method.}
\end{tabular}
\end{center}
\end{table*}

\begin{table*}
\begin{center}
\caption{Observed and bolometric magnitude of the HB red clump, RGB bump and tip for the cluster sample}
\label{bol}
\begin{tabular}{lccccccccccccc}
\hline \hline
Name& [Fe/H] &[M/H] & J$^{RC}$ & H$^{RC}$ & K$^{RC}$ & J$^{bump}$ & H$^{bump}$ & K$^{bump}$ & J$^{tip}$ & H$^{tip}$ & K$^{tip}$ &M$^{bump}_{bol}$ &M$^{tip}_{bol}$  \\
\hline
Ter~1  &  -1.11	&-0.91 &  15.00  &  13.85  &  13.50   &14.60  &  13.60  &  13.10 & 10.356 &   8.913&   8.398&   0.18&    -3.78\\	     
Ter~2  &  -0.72	&-0.53 &  15.00 &  14.00 &  13.60  & 15.23 &  14.30 &  13.80&  10.468&   8.982&   8.511&   0.60&    -3.81\\   
Ter~4  &  -1.58	&-1.37 &  \---  &  \---  &  \---   & \---  &  \---  &  \--- & 10.696 &	 9.333&   8.843&   \---&    -3.57\\
Ter~9  &  -1.21	&-1.01 &  \---  &  \---  &  \---   & \---  &  \---  &  \--- &  9.807 &	 8.549&   8.131&   \---&    -3.86\\
HP~1   &  -1.12	&-0.91 &  \---  &  \---  &  \---   & \---  &  \---  &  \--- &  9.891 &	 8.680&   8.287&   \---&    -3.56\\
Lil~1  &  -0.36	&-0.14 &  16.25 &  14.80 &  14.25  & 17.00 &  15.30 &  14.95&  11.558&   9.624&   8.691&   1.21&    -3.81\\   
Djo~1  &  -1.51	&-1.31 &  \---  &  \---  &  \---   & \---  &  \---  &  \--- & 11.729 & 10.545 & 10.152 &   \---&    -3.68\\
Djo~2  &  -0.65	&-0.45 &  14.10 &  13.40 &  13.15  & 14.40 &  13.75 &  13.40&	9.796&   8.700&   8.168&   0.73&    -3.50\\
NGC~6540  &  -1.29	&-1.10 &  \---  &  \---  &  \---   & \---  &  \---  &  \--- &  8.873 &	 7.903&   7.584&   \---&    -3.56\\
NGC~6544  &  -1.19	&-1.00 &  \---  &  \---  &  \---   & \---  &  \---  &  \--- &  7.600 &	 6.576&   6.242&   \---&    -3.57\\
NGC~6522  &  -1.52	&-1.33 &  \---  &  \---  &  \---   & \---  &  \---  &  \--- &  9.775 &	 8.809&   8.495&   \---&    -3.43\\
NGC~6453  &  -1.57	&-1.38 &  \---  &  \---  &  \---   & \---  &  \---  &  \--- & 10.527 &	 9.577&   9.337&   \---&    -3.57\\
&&&&&&&&&&&&&\\
\hline
\end{tabular}
\end{center}
\end{table*}

\section*{Acknowledgments}

Part of the data analysis was performed with software developed by P. Montegriffo at the Osservatorio Astronomico di Bologna (INAF). The Ministero dell'Istruzione dell'Universit\'a e della Ricerca Italiana is kindly acknowledged for financial support.

The authors warmly thank the ESO La Silla Observatory staff for assistance during the observations. 

This publication makes use of data products from the Two Micron All Sky Survey (2MASS), which is a joint project of the University of Massachusetts and Infrared Processing and Analysis Center/California Institute of Technology, founded by the National Aeronautics and Space Administration and the National Science Foundation.

\label{lastpage}


\begin{thebibliography}{99}
\bibitem[\protect\citeauthoryear{Alcaino}{1983}]{6544_A}
Alcaino, G. 1983, A\&AS, 52,105
\bibitem[\protect\citeauthoryear{Alves\--Brito et al.}{2006}]{alv06}
Alves\--Brito, A., Barbuy, B., Zoccali, M. et al. 2006, A\&A, 460, 269
\bibitem[\protect\citeauthoryear{Armandroff \& Zinn}{1988}]{az88}
Armandroff, T.~E.  \& Zinn, R. 1988, AJ, 96, 92
\bibitem[\protect\citeauthoryear{Barbuy et al.}{1998}]{barb98}
Barbuy, B.,  Bica, E., Ortolani, S. 1998, A\&A, 333, 17
\bibitem[\protect\citeauthoryear{Barbuy et al.}{2006}]{hp1_A}
Barbuy,B.; Zoccali,M.; Ortolani,S.; Momany,Y.; Minniti,D.; Hill,V.; 
Renzini,A.; Rich,R.M.; Bica,E.; Pasquini,L.; Yadav,R.K.S., 2006, A\&A,449,349
\bibitem[\protect\citeauthoryear{Bica \& Pastoriza}{1983}]{bipas83}
Bica, E. \& Pastoriza, M.~G. 1983, Ap\&SS, 91, 99
\bibitem[\protect\citeauthoryear{Bica \& Alloin}{1986}]{bica86}
Bica, E. \& Alloin, D., 1986, A\&A 162, 21
\bibitem[\protect\citeauthoryear{Bica et al.}{1994}]{6540_A}
Bica,E.; Ortolani,S.; Barbuy,B. 1994, A\&A,283,67
\bibitem[\protect\citeauthoryear{Bica et al.}{1998}]{bica98}
Bica, E., Claria, J.~J., Piatti, A.~E., Bonatto, C., 1998, A\$AS, 131, 483
\bibitem[\protect\citeauthoryear{Christian \& Friel}{1992}]{ter2_B}
Christian, C.~A.; Friel,E.~D. 1992,AJ,103,142
\bibitem[\protect\citeauthoryear{Castro et al.}{1995}]{6522_D}
Castro, S.; Barbuy, B.; Bica, E.; Ortolani, S.; Renzini, A. 1995,A\&AS,111,17
\bibitem[\protect\citeauthoryear{Barbuy et al.}{1994}]{6522_E}
Barbuy, B.; Ortolani, S.; Bica, E. 1994,A\&A,285,871

\bibitem[\protect\citeauthoryear{Carretta \& Gratton}{1997}]{cg97}
Carretta, E., \& Gratton, R. G. 1997, A\&AS, 121, 95
\bibitem[\protect\citeauthoryear{Carretta et al.}{2007}]{eug07}
Carretta, E., Bragaglia, A., Gratton, R. G., Momany, Y.; Recio-Blanco, A.; Cassisi, S.; Franois, P.; James, G.; Lucatello, S.; Moehler, S. 2007, A\&A, 464, 967

\bibitem[\protect\citeauthoryear{Cunha \& Smith}{2006}]{katia06}
Cunha, K. \& Smith, V.~V. 2006, ApJ, 651, 491
\bibitem[\protect\citeauthoryear{Davidge}{2000a}]{hp1lil1djo1}
Davidge,T.J.,2000,ApJS, 126,105
\bibitem[\protect\citeauthoryear{Davidge}{2000b}]{dav00}
Davidge,T.J.,2000,AJ, 120,1853
\bibitem[\protect\citeauthoryear{Davidge}{2001}]{dav01}
Davidge,T.J.,2001,AJ, 121,3100


\bibitem[\protect\citeauthoryear{Eisenhauer et al.}{2003}]{eis03}
Eisenhauer, F., Schodel, R., Genzel, R., Ott, T, Tezca, M., Abuter, R., Eckart. A., Alexander, T. 2003, ApJ, 597, L121
\bibitem[\protect\citeauthoryear{Ferraro et al.}{1999}]{frf99}
Ferraro, F. R., Messineo, M., Fusi Pecci, F., De Paolo, M. A., Straniero, O.,
Chieffi, A., \& Limongi, M. 1999, AJ, 118, 1738 
\bibitem[\protect\citeauthoryear{Ferraro et al.}{2000}]{frf00}
Ferraro, F.~R. , Montegriffo, P., Origlia L., Fusi Pecci F.  2000, AJ, 119, 1282 (F00)
\bibitem[\protect\citeauthoryear{Ferraro, Valenti \& Origlia}{2006}]{fvo06}
Ferraro, F.~R., Valenti, E. \& Origlia, L. 2006, ApJ, 649, 243 (FVO06)

\bibitem[\protect\citeauthoryear{Frogel et al.}{1995}]{lil1_D}
Frogel,J.~A.; Kuchinski,L.~E..; Tiede,G.~P., 1995,AJ,109,1154

\bibitem[\protect\citeauthoryear{Fulbright, McWilliam \& Rich}{2007}]{fmr07}
Fulbright, J.~P., McWilliam, A. \& Rich, R.~M. 2007, ApJ, 661, 1152
\bibitem[\protect\citeauthoryear{Glass \& Schultheis}{2003}]{6522_A}
Glass,I.S.; Schultheis,M. 2003,MNRAS,345,39

\bibitem[\protect\citeauthoryear{Harris}{1996}]{har}
Harris, W.~E. 1996, AJ, 112, 1487 (for the updated version see http://physwww.mcmaster.ca/$\sim$harris/mwgc.dat)

\bibitem[\protect\citeauthoryear{Idiart et al.}{2002}]{ter1_A}
Idiart, T.~P., Barbuy, B., Perrin, M.~N., Ortolani, S., Bica, E., Renzini, A. 2002, A\&A, 381, 472

\bibitem[\protect\citeauthoryear{Minniti, Oleszewski \& Rieke}{1995}]{min95}
Minniti,D. 1995, AJ, 110,1686
\bibitem[\protect\citeauthoryear{Lecureur et al.}{2007}]{lec07}
Lecureur, A., Hill, V., Zoccali, M., Barbuy, B., Gomez, A., Minniti, D., Ortolani, S., Renzini, A. 2007, A\&A, 465, 799L

\bibitem[\protect\citeauthoryear{Montegriffo et al.}{1998}]{bol98}
Montegriffo, P., Ferraro, F.~R., Origlia, L., Fusi Pecci, F. 1998, MNRAS, 297, 872
\bibitem[\protect\citeauthoryear{Origlia et al.}{2005}]{ori05}
Origlia, L., Valenti, E., Rich, R.~M., Ferraro, F.~R. 2005, MNRAS, 363, 897
\bibitem[\protect\citeauthoryear{Origlia et al.}{2001}]{lil1_B}
Origlia,L.; Rich,R.M.; Castro,S.M. 2001,A\&AS, 199,5616
\bibitem[\protect\citeauthoryear{Origlia \& Rich}{2004}]{ter4_B}
Origlia, L.; Rich,R.~M. 2004,AJ,127,3422
\bibitem[\protect\citeauthoryear{Origlia et al.}{2008}]{ori08}
Origlia,L.; Valenti, E. \& Rich,R.M. 2008, MNRAS, 388, 1419

\bibitem[\protect\citeauthoryear{Ortolani et al.}{1995a}]{djo1_A}
Ortolani,S.; Bica,E.; Barbuy,B. 1995,A\&A,296,680
\bibitem[\protect\citeauthoryear{Ortolani et al.}{1995b}]{ter96453}
Ortolani,S.; Bica,E.; Barbuy,B. 1999,A\&AS,138,267
\bibitem[\protect\citeauthoryear{Ortolani et al.}{1996}]{lil1_C}
Ortolani, S.; Bica,E.; Barbuy,B. 1996,\&A,306,134
\bibitem[\protect\citeauthoryear{Ortolani et al.}{1997a}]{djo2_A}
Ortolani,S.; Bica,E.; Barbuy,B. 1997, A\&AS,126,319
\bibitem[\protect\citeauthoryear{Ortolani et al.}{1997b}]{hp1_B}
Ortolani,S.; Bica,E.; Barbuy,B. 1997, MNRAS, 284,692
\bibitem[\protect\citeauthoryear{Ortolani et al.}{1997c}]{ter4_C}
Ortolani, S.; Barbuy, B.; Bica, E. 1997, A\&A, 319,850
\bibitem[\protect\citeauthoryear{Ortolani et al.}{1999a}]{ter1_B}
Ortolani, S., Barbuuy, B., Bica, E., Renzini, A., Marconi, G., Gilmozzi, R. 1999, A\&A, 350, 840
\bibitem[\protect\citeauthoryear{Ortolani et al.}{1999b}]{ter2_A}
Ortolani,S.; Bica,E.; Barbuy,B. 1997, A\&A, 326, 614
\bibitem[\protect\citeauthoryear{Ortolani et al.}{2007a}]{lil1ter4_A}
Ortolani,S.; Barbuy,B.; Bica,E.; Zoccali,M.; Renzini,A. 2007,A\&A,470,1043


\bibitem[\protect\citeauthoryear{Rich \& Origlia}{2005}]{ro05}
Rich, R.~M. \& Origlia, L. 2005, ApJ, 634, 1293
\bibitem[\protect\citeauthoryear{Rich et al.}{2007}]{rov07}
Rich, R.~M., Origlia, L. \& Valenti, E. 2007, ApJ, 665L, 119

\bibitem[\protect\citeauthoryear{Stetson}{1987}]{dao}
Stetson, P.~B. 1987, PASP, 99, 191
\bibitem[\protect\citeauthoryear{Shara et al.}{1998}]{6522_C}
Shara,M.M.; Drissen,L.; Rich,R.M.; Paresce,F.; King,I.R.; Meylan,G, 1998,ApJ,495,796
\bibitem[\protect\citeauthoryear{Stephens \& Frogel}{2004}]{ter4lil1ter2}
Stephens, A.~W.; Frogel, J.~A.. 2004,AJ,127,925

\bibitem[\protect\citeauthoryear{Terndrup et al.}{1998}]{6522_B}
Terndrup,D.~M.; Popowski,P.; Gould,A.;Rich,R.M.; Sadler,E.~M. 1998,AJ,115,1476
\bibitem[\protect\citeauthoryear{Trager et al.}{1993}]{tra93}
Trager, S.~C., Djorgovski, S., King, I~R., 1993, in: Djorgovski, S., Meylan, G. (eds) Structure and Dynamics of Globular Clusters, ASP Conference Sr.
\bibitem[\protect\citeauthoryear{Valenti et al.}{2004}]{io04}
Valenti, E., Ferraro, F.~R. , Perina, S., Origlia L.  2004a, A\&A, 419, 139
\bibitem[\protect\citeauthoryear{Valenti, Ferraro \& Origlia}{2004a}]{io04a}
Valenti, E., Ferraro, F.~R. \& Origlia L. 2004a, MNRAS, 351, 1204 (VFO04a)
\bibitem[\protect\citeauthoryear{Valenti, Ferraro \& Origlia}{2004b}]{io04b}
Valenti, E., Ferraro, F.~R. \& Origlia L. 2004b, MNRAS, 354, 815 (VFO04b)
\bibitem[\protect\citeauthoryear{Valenti, Origlia \& Ferraro}{2005}]{io05}
Valenti, E. , Origlia L.  \& Ferraro, F.~R.2005, MNRAS, 361, 272
\bibitem[\protect\citeauthoryear{Valenti, Ferraro \& Origlia}{2007}]{io07}
Valenti, E., Ferraro, F.~R. \& Origlia L. 2007, AJ, 133, 1287 ({\it Paper~I})
\bibitem[\protect\citeauthoryear{Zoccali et al.}{2004}]{zoc04}
Zoccali, M., Barbuuy, B., Hill, V., Ortolani, S., Renzini, A., Bica, E., Momany, Y., Pasquini, L., Minniti, D., Rich, R.~M., 2004, A\&A, 423, 507
\bibitem[\protect\citeauthoryear{Zoccali et al.}{2006}]{zoc06}
Zoccali, M., Lecureur,, A., Barbuy, B., Hill, V., Renzini, A., Minniti, D., Momany, Y., Gomez., A., Ortolani, S., 2006, A\&A, 457L, 1
\bibitem[\protect\citeauthoryear{Zoccali et al.}{2008}]{zoc08}
Zoccali, M., Hill, V., Lecureur, A., Barbuy, B., , Renzini, A., Minniti, D., Gomez., A., Ortolani, S., 2008, A\&A, 486, 177
\bibitem[\protect\citeauthoryear{Zinn \& West}{1984}]{zw84}
Zinn, R., \& West, M.~J., 1984, ApJS 55, 45
\bibitem[\protect\citeauthoryear{Zinn}{1985}]{zin85}
Zinn, R. 1985, ApJ, 293, 424
\bibitem[\protect\citeauthoryear{Weiland et al.}{1994}]{wei94}
Weiland, J.~L., et al. 1994, ApJ, 425, L81
\end{thebibliography}
\end{document}